\documentclass[11pt]{article}
\usepackage{cite}
\usepackage{amsmath,amsfonts,amssymb}
\usepackage[small,bf,hang]{caption}
\usepackage{slashed}
\usepackage[vcentermath]{youngtab}

\def\hybrid{
        \topmargin -20pt
        \oddsidemargin 0pt
        \headheight 0pt \headsep 0pt
        \textwidth 6.25in 
        \textheight 9.5in 
        \marginparwidth .875in
        \parskip 5pt plus 1pt \jot = 1.5ex}

\hybrid
\usepackage{tikz}

 \csname
@addtoreset\endcsname{equation}{section}

\usepackage{color}
\definecolor{red}{rgb}{1,0,0}
\definecolor{lred}{rgb}{0.3,0,0}
\definecolor{green}{rgb}{0,0.6,0}
\definecolor{blue}{rgb}{0,0,1}
\definecolor{violet}{rgb}{0.8,0,0.8}
\definecolor{amber}{rgb}{1.0, 0.75, 0.0}
\definecolor{yellow}{rgb}{1.0, 1.0, 0.0}
\definecolor{applegreen}{rgb}{0.55, 0.71, 0.0}
\definecolor{cadmiumgreen}{rgb}{0.0, 0.42, 0.24}
\definecolor{ballblue}{rgb}{0.13, 0.67, 0.8}
\definecolor{caribbeangreen}{rgb}{0.0, 0.8, 0.6}
\definecolor{bluemunsell}{rgb}{0.0, 0.5, 0.69}
\definecolor{brightpink}{rgb}{1.0, 0.0, 0.5}

\newcommand{\D}{\mathcal{D}}
\newcommand{\M}{\mathcal{M}}
\newcommand{\J}{\mathcal{J}}
\newcommand{\K}{\mathcal{K}}
\newcommand{\Z}{\mathcal{Z}}
\renewcommand{\L}{\mathcal{L}}
\newcommand{\T}{\mathcal{T}}

\newcommand{\Ld}{\mathbb{L}}
\newcommand{\Dd}{\mathbb{D}}
\newcommand{\E}{\mathbb{E}}

\newcommand{\Cp}{f_1{}}
\newcommand{\Cm}{f_2{}}

\newcommand{\nn}{\nonumber}

\newcommand{\qq}{\quad , \quad}

\newcommand{\Tr}[1]{\text{Tr}\left(#1\right)}

\def\moth{\mathsurround=0pt}
\newdimen\zo \zo=0pt

\def\tick{\leaders\hrule height 0.5ex depth 0pt \hskip 0.5pt}
\def\upboxfill{$\moth \setbox\zo\hbox{\tick}%
  \hskip 3pt\hbox to 0pt{$\tick$\hss}\hrulefill \hbox to 7.5pt{$\tick$\hss}$}

\def\dtick{\leaders\hrule height .34pt depth 0.5ex \hskip 0.5pt}
\def\downboxfill{$\moth \setbox\zo\hbox{\dtick}%
  \hskip 2pt\hbox to 0pt{$\dtick$\hss}\hrulefill \hbox to 2pt{$\dtick$\hss}$}

\def\bec{\begin{center}}
\def\ec{\end{center}}

\def\D{{\cal D}}

\def\qq{\quad\quad}

\def\nn{\nonumber}

\def\Tr{{\rm Tr}}

\def\be{\begin{equation}}
\def\ee{\end{equation}}
\def\bea{\begin{eqnarray}}
\def\eea{\end{eqnarray}}
\def\ba{\begin{array}}
\def\ea{\end{array}}

\begin{document}

\begin{titlepage}
	\rightline{}
	\rightline{November 2022}
	\rightline{HU-EP-22/34-RTG} 
	\begin{center}
		\vskip 1.5cm
		 
		 {\Large \bf{ An $\alpha'$-complete theory of cosmology  \\ [1ex]
		 and its tensionless limit }}

		\vskip 1.7cm

		{\large\bf {Tomas Codina$^\dag$, Olaf Hohm$^\dag$ and Diego Marques$^*$}}
		\vskip 1cm
		
		$^\dag$ {\it   Institute for Physics, Humboldt University Berlin,\\
			Zum Gro\ss en Windkanal 6, D-12489 Berlin, Germany}\\
		
		\vskip .3cm
		
		$^*$ {\it   Instituto de Astronom\'ia y F\'isica del Espacio, \\
			Casilla de Correo 67 - Suc. 28 (C1428ZAA), Buenos Aires, Argentina}\\
		\vskip .1cm
		
		\vskip .4cm

		tomas.codina@physik.hu-berlin.de, ohohm@physik.hu-berlin.de, diegomarques@iafe.uba.ar

		\vskip .4cm

	\end{center}

	\bigskip\bigskip
	\begin{center} 
		\textbf{Abstract}
		
	\end{center} 
	\begin{quote}  
		
		We explore the exactly duality 
		invariant higher-derivative extension of double field theory 
		due to Hohm, Siegel and Zwiebach (HSZ) specialized to cosmological backgrounds. 
		Despite featuring a finite number of derivatives in its original formulation, 
		this theory encodes infinitely many  $\alpha'$ corrections 
		for metric, B-field and dilaton, which are obtained upon integrating out certain extra fields. 
		We perform a cosmological reduction with fields  depending  only on time and show consistency 
		of this truncation. We compute the $\alpha'^4$ coefficients of the general cosmological classification. 
		As a possible model for how to deal with all  $\alpha'$ corrections in string theory 
		we  give  a two-derivative reformulation in which the extra fields are kept. 
		The corresponding Friedmann equations are then 
		ordinary second order differential equations  that  capture all $\alpha'$ corrections. 
		We explore the tensionless  limit $\alpha'\rightarrow \infty$, which features
		 string frame de Sitter vacua, and we set up perturbation theory in $\frac{1}{\alpha'}$. 
		
	\end{quote} 
	\vfill
	\setcounter{footnote}{0}
\end{titlepage}

\newpage

\tableofcontents

\section{Introduction}

A basic  obstacle for understanding string theory, let alone  confronting it  with observation,  is that 
we do not know the defining equations of string theory. We know the low-energy effective theories given by (super)gravity, 
but even classical string theory restricted to the massless fields features an infinite number of 
higher-derivative $\alpha'$ corrections going beyond  conventional gravity. 
While these corrections have been determined to a few orders in $\alpha'$, 
a computation of all $\alpha'$ corrections is out of reach. 
But it is precisely in regimes where a gravitational theory beyond general relativity is almost certainly needed 
(singularities of cosmology or black holes), that one plausibly expects 
\textit{all} $\alpha'$ corrections to be important. (See 
\cite{Brandenberger:1988aj,Tseytlin:1991xk,Veneziano:1991ek,Gasperini:1996fu,Gasperini:2002bn,Bernardo:2020nol,Bernardo:2020bpa,Nunez:2020hxx}  
for classic and more recent string cosmology proposals 
requiring  the inclusion of all $\alpha'$ corrections.)

In this paper we explore a particular spacetime theory 
based on double field theory \cite{Siegel:1993th, Hull:2009mi, Hohm:2010pp, Hohm:2013bwa}, 
due to Hohm, Siegel and Zwiebach (HSZ)  \cite{Hohm:2013jaa}, and apply it to cosmological backgrounds. 
Our goal is two-fold, namely first to  compute the $\alpha'^4$ coefficients of the general cosmological classification, 
thereby  going beyond the current 
state of the art \cite{Codina:2020kvj,Codina:2021cxh,Bonezzi:2021sih}, 
and second and 
perhaps more intriguingly,  to have a model for a theory that is `$\alpha'$-complete' in  a sense made precise below.   
The construction of HSZ theory was based on a non-standard chiral CFT and is thus not a conventional 
string theory. (It also appears  closely related to the `chiral string theory' of \cite{Huang:2016bdd}
and/or to the ambitwistor string \cite{Casali:2016atr}, but to our knowledge the precise connections 
have  not been established.) 
Nevertheless, HSZ theory shares crucial features of any string theory, such as 1) duality invariance under 
$O(d,d,\mathbb{R})$ for backgrounds with $d$ Abelian isometries, and 2) 
the presence of infinitely many higher-derivative corrections 
for the massless fields. While in its original formulation 
HSZ theory carries only up to six derivatives, it also features extra massive fields, 
in addition to the familiar massless fields of string theory (metric, B-field and dilaton), and integrating out these extra fields induces 
an infinite tower of  $\alpha'$ corrections for the massless fields that are kept. These higher-derivative 
corrections include a Green-Schwarz-type deformation at order $\alpha'$ and a Riemann-cube invariant 
at order $\alpha'^2$ \cite{Hohm:2014eba,Hohm:2014xsa,Naseer:2016izx}, but beyond that only very little is known.

In order to understand  the theory in a simplified setting, and to relate to the physically interesting  case  of cosmology, 
we specialize HSZ  theory to cosmological backgrounds, i.e.~with fields depending only on time. 
We then show how to integrate out the extra fields so as to compute the order 
$\alpha'^4$ coefficients of the general cosmological classification \cite{Hohm:2015doa,Hohm:2019jgu}. 
As a second main result of this paper, we then show  that, rather than integrating out the extra fields, 
we can keep them  and reformulate the theory so that it features only up to two derivatives. 
Such a reformulation was given in \cite{Hohm:2016lim} for HSZ theory without  reduction, but restricted  to  quadratic order in a background field expansion  
about flat space.  
Here we exhibit the corresponding  reformulation in the cosmological reduction but including all non-linear 
couplings. 
Intriguingly, in this formulation the tensionless limit $\alpha'\rightarrow \infty$ can be taken smoothly. 
We also present the equations of motion for Friedmann-Robertson-Walker backgrounds, 
which  by construction capture all $\alpha'$ corrections.

It has sometimes  been speculated that the tensionless limit of string theory, if it were accessible,  
would be a powerful tool  for learning about  the fundamental degrees of freedom of string theory \cite{Gross:1988ue, Atick:1988si}. 
We thus take the opportunity and  explore the cosmological Friedmann equations in the tensionless limit, in which 
we find the most general solutions. Remarkably, these solutions include string frame de Sitter vacua 
(that are also de Sitter vacua in Einstein frame for $d=4$, i.e., in five spacetime dimensions). 
Interestingly,  the de Sitter length scale here is not a bare parameter in the action but rather 
appears as an integration constant. 
In order to find solutions of the full equations we then use perturbation theory, but not in $\alpha'$ as would 
be appropriate for small $\alpha'$, but rather in $\frac{1}{\alpha'}$ as is appropriate for 
perturbations around infinite $\alpha'$. We find that the de Sitter vacua survive the first order corrections in $\frac{1}{\alpha'}$.

We close  this introduction by discussing the 
$\alpha'$-completeness of this theory. Gravity theories  with a finite number of higher derivatives  generically 
suffer from various pathologies, as the presence of unphysical ghost modes 
in the spectrum.\footnote{Counter examples are the Starobinsky model \cite{Starobinsky:1980te}, 
	which augments the Einstein-Hilbert term 
	by the square of the Ricci scalar, and `new massive gravity' \cite{Bergshoeff:2009hq}, with  a particular curvature-squared modification 
	of 3D gravity.} 
For this reason, the usual view is that higher-derivative modifications only make sense 
in perturbation theory, with features of the two-derivative theory being corrected by a small parameter 
like $\alpha'$ and carrying 
an  infinite number of such higher-derivative corrections. Another perspective showing that a finite number of higher-derivative corrections is problematic is 
that of $O(d,d,\mathbb{R})$ invariance for backgrounds with $d$ Abelian isometries, in the following referred 
to as `duality', which string theory must possess to all orders in $\alpha'$ \cite{Sen:1991zi}. 
While in the two-derivative truncation this duality is easily recognized and realized exactly, once higher-derivative terms are included the 
situation becomes more subtle \cite{Meissner:1996sa}. 
Adding the known four-derivative (order $\alpha'$) 
terms in the bosonic string action 
and performing dimensional reduction along $d$ dimensions, the resulting theory does \textit{not} possess 
the expected duality invariance. It is, however, invariant under  higher-derivative 
deformations of the duality transformations \textit{up to terms of order $\alpha'^2$} \cite{Meissner:1996sa}. Thus, 
within  a theory with infinitely many $\alpha'$ corrections this shows compatibility with duality,  
but a finite higher-derivative modification is generally incompatible with duality. 
The HSZ theory to be investigated here is exactly duality invariant and in this sense $\alpha'$-complete. 

The rest of this paper is organized as follows. 
In sec.~2 we introduce the cosmological ansatz for HSZ theory, which is non-trivial in that certain off-diagonal 
field components need to be kept. We prove consistency of this  truncation and give the cosmological action. 
This is used in sec.~3 in order to integrate out the extra fields up to and including order $\alpha'^4$. 
The resulting action is then brought to the canonical form of the cosmological classification \cite{Hohm:2019jgu} by means of 
field redefinitions. In sec.~4 we present  a reformulation of the cosmological HSZ action, including the extra fields, 
with only two derivatives. In sec.~5 we give the corresponding $\alpha'$-complete Friedmann equations, take the tensionless limit and find the most general solutions which include a de Sitter vacua as a particular case. We finish this section by computing the first $\frac{1}{\alpha'}$ correction to this specific solution. We conclude with a brief outlook in sec.~6.

\section{Consistent truncation to cosmological backgrounds}

In this section we review the double $\alpha'$-geometry by Hohm, Siegel and Zwiebach \cite{Hohm:2013jaa}, here referred to as HSZ theory. We will introduce the field content, symmetries and action. We then perform the truncation that will take us to the time-dependent cosmological action in terms of (duality covariant) scalar fields.  

\subsection{Review of HSZ theory}

The field content of HSZ theory includes  the `double metric'  $\M_{\hat{M} \hat{N}}$, 
with $O(D,D)$ indices $\hat{M}, \hat{N} = 1, \dots, 2D$, and a generalized dilaton field $d$. They are defined on a double space with coordinates $X^{\hat M}$, but  all products of fields and gauge parameters are annihilated by the duality invariant Laplacian 
\begin{equation}\label{SC}
\eta^{\hat M \hat N} \partial_{\hat M} \partial_{\hat N} \cdots = 0\;,
\end{equation}
where $\eta_{\hat M \hat N}$ is the $O(D,D)$ invariant metric that raises and lowers indices. This is known as the strong constraint. 
In contrast to  the standard generalized metric formulation of double field theory (DFT)\cite{Siegel:1993th, Hull:2009mi, Hohm:2010pp, Hohm:2013bwa}, where the generalized metric ${\cal H}_{\hat M \hat N}$ obeys the constraint 
${\cal H}\eta^{-1}{\cal H}=\eta$, the double metric is symmetric but otherwise unconstrained. 
Under infinitesimal generalized diffeomorphisms with parameter $\xi^{\hat{M}}$, the generalized dilaton transforms as in  conventional DFT
\begin{equation}\label{HSZdtransform}
\delta d = \xi^{\hat{P}} \partial_{\hat{P}} d - \frac12 \partial_{\hat{P}} \xi^{\hat{P}}\,, 
\end{equation}
while the double metric receives linear and quadratic corrections in $\alpha'$
\begin{equation}\label{HSZ_transformations}
\begin{aligned}
\delta \M_{\hat{M} \hat{N}} = 
\widehat{\mathcal{L}}_{\xi} \M_{\hat{M} \hat{N}} + \alpha' \J^{(1)}_{\hat{M} \hat{N}} + \alpha'{}^2 \J^{(2)}_{\hat{M} \hat{N}}\,, 
\end{aligned} 
\end{equation} 
where the standard generalized Lie derivative is defined by 
\begin{equation}
\widehat{\mathcal{L}}_{\xi} \M_{\hat{M} \hat{N}} \equiv \xi^{\hat{P}} \partial_{\hat{P}} \M_{\hat{M} \hat{N}} + \K_{\hat{M}}{}^{\hat{P}} \M_{\hat{P} \hat{N}} + \K_{\hat{N}}{}^{\hat{P}} \M_{\hat{M} \hat{P}}\;, \qquad
\K_{\hat{M} \hat{N}} \equiv 2 \partial_{[\hat{M}} \xi_{\hat{N}]}\;, 
\end{equation} 
and the higher-derivative contributions are given by 
\begin{equation}\label{HSZ_J}
\begin{aligned}
\J^{(1)}_{\hat{M} \hat{N}} &\equiv - \frac12 \partial_{\hat{M}} \M^{\hat{P} \hat{Q}} \partial_{\hat{P}} \K_{\hat{Q} \hat{N}} - \partial_{\hat{P}} \M_{\hat{Q} \hat{M}} \partial_{\hat{N}} \K^{\hat{Q} \hat{P}} + (\hat{M} \rightleftarrows \hat{N})\;,\\
\J^{(2)}_{\hat{M} \hat{N}} &\equiv - \frac14 \partial_{\hat{M} \hat{K}} \M^{\hat{P} \hat{Q}} \partial_{\hat{N} \hat{P}} \K_{\hat{Q}}{}^{\hat{K}} + (\hat{M} \rightleftarrows \hat{N}) \;. 
\end{aligned}
\end{equation}
These transformations close under a deformation of the C-bracket
\begin{equation}
\big[\xi_1, \xi_2\big]^{\hat{M}}_{(C)} \equiv 2 \xi^{\hat{P}}_{[1}\partial_{\hat{P}} \xi^{\hat{M}}_{2]} - \xi^{\hat{P}}_{[1}\partial^{\hat{M}}\xi_{2] \hat{P}} \,,
\end{equation} 
 which we call the $C'$-bracket: 
\begin{equation}\label{C'}
\begin{aligned}
\big[\xi_1, \xi_2\big]^{\hat{M}}_{(C')} \equiv \big[\xi_1, \xi_2\big]^{\hat{M}}_{(C)} +\alpha' 
\partial_{\hat{P}} \xi^{\hat{Q}}_{[1} \partial^{\hat{M}} \partial_{\hat{Q}} \xi^{\hat{P}}_{2]}\,. 
\end{aligned}
\end{equation}
The dynamics of the theory is encoded in an action, that can be written compactly as
\begin{equation}\label{IHSZ}
I_{\rm HSZ} = \int d^{2 D} \, X \, e^{-2 d} \, \left\langle \M \,  \left| \, \eta - \frac16 \M \star \M \right.\right\rangle\;.
\end{equation}
The definitions for the inner product $\left\langle \, \cdot \, | \, \cdot \, \right\rangle$ and star-product $\star$ involve long expressions in terms of $\M$, $d$ and combinations of them up to and including six derivatives. Since these explicit expressions are not very illuminating, we refer the reader to equations (2.11), (2.12) and (2.14) of \cite{Lescano:2016grn} where the definitions are given in detail, or to appendix D therein where the complete action is given. The exact gauge symmetry under \eqref{HSZdtransform} and \eqref{HSZ_transformations} can be checked once the definitions of the products are used, together with the strong constraint \eqref{SC}.

\subsection{The cosmological truncation}

We will now truncate the above theory to a cosmological ansatz in which fields depend only on time, 
and  we will prove that the truncation is consistent. 
To this end we assume a split of the coordinates and indices as follows
\begin{equation}
X^{\hat{M}} = \big(t\,,\; \tilde t\,, \;Y^M\big) \;, \qquad   {}_{\hat{M}} = \big(_0\,,\;  {}^0\,, \; _M\big) \;, \qquad M = 1, \dots, 2(D-1)\;. 
\end{equation}
This ansatz  breaks the manifest  $O(D, D)$ invariance to $O(1, 1) \times O(D-1, D-1)$. Furthermore, we will solve the strong constraint by selecting a frame in which the fields do not depend on $\tilde t$ nor $Y^M$, 
which breaks the $O(1,1)$ factor and importantly \textit{preserves} the internal $O(D-1, D-1)$. We will thus set 
\begin{equation}
\partial^0 = \partial_M  = 0\;,
\end{equation}
everywhere in the field equations and gauge transformations. 

Let us now turn to the decompositions of the fields and the $O(D,D)$ metric, which are given by 
\begin{equation}\label{ansatz}
\eta_{\hat{M} \hat{N}} = \begin{pmatrix}
0 & 1 & 0\\
1 & 0 & 0\\
0 & 0 & \eta_{M N}
\end{pmatrix} , \qquad  \M_{\hat{M} \hat{N}} = \begin{pmatrix}
- n^{2} B & A & 0\\
A & - \frac{1}{n^2} & 0\\
0 & 0 & \Z_{M N}
\end{pmatrix}  , \qquad d = \frac12 \Phi - \frac12 \ln n\;, 
\end{equation}
where all  fields depend only on time $t$. Apart from the exclusive time dependence, the only truncation applied above is given by the vanishing of some components of the double metric. We explain latter why setting these `vector' components to zero is a consistent choice. Additionally, one would be tempted to further reduce the external $2\times 2$ block in the double metric to be diagonal and $O(1,1)$ valued by setting $A = 0$, $B = 1$, but this turns out to be inconsistent. By inspecting the action of generalized diffeomorphisms  (\ref{HSZ_transformations}) for this ansatz, with the gauge parameter decomposed as 
\begin{equation}\label{xiansatz}
\xi^{\hat{M}} =  \big(\xi^0, \xi_0,\xi^M \big) \equiv \big(\xi, 0, { 0}\big)\;, 
\end{equation}
we find the following gauge transformations for the component fields 
\begin{equation}\label{cosmoHSZ_transformations}
\begin{aligned}
\delta n &= \xi \dot{n} + n \dot{\xi}\,, \\
\delta A &= \xi \dot{A} - 3 \alpha' \ddot{\xi} \frac{\dot{n}}{n^3}\,,\\
\delta B &= \xi \dot{B} + \alpha' \frac{1}{n^2} \ddot{\xi} \dot{A} + \alpha'{}^2 \frac{1}{n^5} \dddot{\xi} \left( \ddot{n} - 3 \frac{\dot{n}^2}{n}\right)\,,\\
\delta \Z_{M N} &= \xi \dot{\Z}_{M N}\,, \\
\delta \Phi &= \xi \dot{\Phi}\, ,
\end{aligned}
\end{equation}
where here and in the following the dot denotes 
time derivatives, i.e.~$\partial_t \Psi  = \partial_0 \Psi \equiv \dot{\Psi}$.  
To zeroth order in $\alpha'$ we recognize the familiar transformations under time reparametrizations 
$t\rightarrow t-\xi(t)$, but these transformations receive $\alpha'$-corrections. The fact that the corrections to the transformations of $A$ and $B$ contain corrections not depending on $A$ and $B$ themselves, prevents us from setting them to a constant, so both $A$ and $B$ {\it must} be kept for consistency of the truncation.

Instead, setting to zero the vectorial components of the double metric ${\cal M}_{0 M} = {\cal M}{}^0{}_M = 0$ is perfectly consistent. In fact, the  transformations (\ref{xiansatz}) acting on these components vanish when the components themselves are set to zero, contrary to what happens with $A$ and $B$. Moreover, and equally important, the equations of motion of these components also vanish when the components do. Put differently, the information contained in the equations of motion of all fields after setting ${\cal M}_{0 M} = {\cal M}{}^0{}_M = 0$, is the same as the one obtained by setting these components to zero in the action and then computing the equations of motion of the fields we kept. As a consequence of this truncation, the full original gauge symmetry of HSZ (\ref{HSZ_transformations}) is now broken to time reparametrizations (\ref{xiansatz}).

It is instructive  to inspect also the gauge algebra under this cosmological reduction.  
One obtains with \eqref{C'} 
\begin{align}
\big[\xi_1, \xi_2\big]_{0 \, (C')} & = 2 \xi_{[1} \dot{\xi}_{2]}\;, \\
\big[\xi_1, \xi_2\big]^0_{(C')} & = \alpha' \dot{\xi}_{[1} \ddot{\xi}_{2]} = \left[ \sqrt{\tfrac{\alpha'}{2}}\dot{\xi}_1, \sqrt{\tfrac{\alpha'}{2}}\dot{\xi}_2\right]_{0 \, (C')}\;, \\
\big[\xi_1, \xi_2\big]^M_{(C')} & =  0 \;. 
\end{align}
Given the relationship in the second line, one may suspect that  the only surviving algebra is that of standard one-dimensional diffeomorphisms, suggesting that there  should be a field basis in which this symmetry is realized in the standard way. Indeed, we can find an explicit field redefinition that removes the higher-derivative terms in $\delta A$ and $\delta B$. 
To this end, it is convenient to introduce the derivative operator 
 \begin{equation}
  \D \equiv \frac1n \partial_t\,, 
 \end{equation}
which is covariant under conventional time reparametrizations. 
Specifically, writing the original fields in terms of new primed fields as 
\begin{equation}\label{redefinitions}
\begin{aligned}
A  &= A' - \frac32 \alpha' (\D\ln n)^2\;,\\
B &= B' + \alpha' (\D \ln n) \D A' - \alpha'{}^2 \left[\frac14 (\D \ln n)^4 + (\D^2 \ln n) (\D \ln n)^2 - \frac12 (\D^2 \ln n)^2\right]\,, 
\end{aligned}
\end{equation}
it is straightforward to verify that for $A'$ and $B'$ being reparametrization scalars,  the higher-derivative terms 
induce precisely the higher-order corrections in (\ref{cosmoHSZ_transformations}). 
Furthermore, it is immediate  that the above relations can be inverted hence proving that this is a legal field redefinition.  
All in all, we can express the theory in terms of fields given by  the lapse function $n$ and a number of 
reparametrization scalars, with transformation rules 
\begin{equation}
\delta n = \xi \dot{n} + n \dot{\xi}\,, \ \ \ 
\delta A = \xi \dot{A}\,, \ \ \
\delta B = \xi \dot{B}\,, \ \ \
\delta \Z_{M N} = \xi \dot{\Z}_{M N}\,, \ \ \
\delta \Phi = \xi \dot{\Phi}\,, 
\end{equation}
where we removed the primes from $A'$ and $B'$.

\subsection{The cosmological action}

We now give the HSZ action in this cosmological reduction, which is obtained by plugging the ansatz \eqref{ansatz} together with the field redefinitions \eqref{redefinitions} into \eqref{IHSZ}. More precisely, we performed the reduction at the level of the inner and star products using their explicit definitions as given in \cite{Lescano:2016grn} and then combined these results back into the form of the action \eqref{IHSZ}. As a consistency check we can use that in this field basis the diffeomorphisms act in the usual way, which implies that the derivatives of the lapse function should combine to form covariant derivatives $\D = \frac1n \partial_t$ of the scalar fields.\footnote{As a second consistency check, we also performed the reduction directly at the level of the $\alpha'$-expanded action as given in Appendix D of \cite{Lescano:2016grn}. We used Cadabra2 \cite{Peeters:2007wn} for the reductions and consistency checks.}  
We find that the final manifestly gauge invariant action is given by \begin{equation}\label{cosmoHSZ_action}
\begin{aligned}
I &= \int dt \, n \, e^{-\Phi} \,\left[\frac{1}{\alpha'} \mathcal{L}^{(-1)} + \mathcal{L}^{(0)} + \alpha' \mathcal{L}^{(1)} + \alpha'{}^2 \mathcal{L}^{(2)}\right]\;,
\end{aligned}
\end{equation}
where
\begin{equation}\label{cosmoHSZ_lagrangian1}
\begin{aligned}
\mathcal{L}^{(-1)} &= A - \frac{1}{3} A^3 - A B + \frac12 \Tr{\Z} -\frac16 \Tr{\Z^3}\,, \\
\mathcal{L}^{(0)} &= -\frac{31}{12} (\D A)^2 - \frac32 A^2 \left[(\D \Phi)^2 - 2 \D^2 \Phi\right] + \frac{35}{6} \D \Phi A \D A - \frac{17}{6} A \D^2 A - \frac{5}{12} \D^2 B + \frac{11}{12} \D \Phi \D B \\
&  - \frac{1}{2} B \left[(\D \Phi)^2 - 2 \D^2 \Phi\right] - \frac{1}{24}\Tr{(\D \Z)^2} + \frac{1}{12} \Tr{\Z \D^2 \Z} -\frac{1}{12} \D \Phi \Tr{\Z \D \Z} + \frac32 (\D \Phi)^2 - 3 \D^2 \Phi\,, \\
\mathcal{L}^{(1)} &= -\frac14 \D^4 A - \frac34 (\D \Phi)^2 \D^2 A + \frac14 (\D \Phi)^3 \D A - 4 (\D^2 \Phi)^2 A + \frac34 \D \Phi \D^3 A + 3 \D^2 \Phi \D^2 A \\
& \quad + \frac92 \D^3 \Phi \D A + \frac52 \D^4 \Phi A   + \frac52 (\D \Phi)^2 \D^2 \Phi A - \frac{11}{2} \D \Phi \D^2 \Phi \D A - 5 \D \Phi \D^3 \Phi A\,, \\
\mathcal{L}^{(2)} &= \frac32 (\D^2 \Phi)^3 - 2 (\D^3 \Phi)^2 + \frac14 \D^6 \Phi - (\D \Phi)^2 (\D^2 \Phi)^2  + \frac34 (\D \Phi)^2 \D^4 \Phi - \frac14 (\D \Phi)^3 \D^3 \Phi\\
& \quad - \frac34 \D \Phi \D^5 \Phi  - \frac{11}{4} \D^2 \Phi \D^4 \Phi + \frac{19}{4} \D \Phi \D^2 \Phi \D^3 \Phi\,. 
\end{aligned}
\end{equation}
By performing integration by parts one may verify  that \eqref{cosmoHSZ_lagrangian1} is equivalent to the simpler action 
of the same structural form (\ref{cosmoHSZ_action}) but with Lagrangian
\begin{equation}\label{cosmoHSZ_lagrangian2}
\begin{aligned}
\mathcal{L}^{(-1)} &= \frac12 \Tr{\Z} -\frac16 \Tr{\Z^3} + A - \frac{1}{3} A^3 - A B\,,\\
\mathcal{L}^{(0)} &= - \frac{1}{8}\Tr{(\D \Z)^2} - \frac32 (\D \Phi)^2 + \frac{1}{4} (\D A)^2 + \frac32 A^2 \D^2 \Phi + \frac12 B \D^2 \Phi\,, 
\\
\mathcal{L}^{(1)} &= \frac12 A \left[ \D^4 \Phi - \D\Phi \D^3 \Phi - 3 (\D^2 \Phi)^2\right]\,, \\
\mathcal{L}^{(2)} &= \frac14 (\D^3 \Phi)^2 + \frac12 (\D^2 \Phi)^3\,. 
\end{aligned}
\end{equation}
Finally, we can bring the full action to its simplest form by performing the following field redefinition
\begin{eqnarray}
B' \!\!\!\! &=& \!\!\!\! B - 1 + \frac13 A^2 -\alpha'\left[\frac43 A \D^2 \Phi + \frac14 \D \Phi \D A - \frac14 \D^2 A\right]  -\frac{\alpha'{}^2}{2}\left[\frac34 \D^4 \Phi - \frac34 \D\Phi \D^3 \Phi -\frac53 (\D^2\Phi)^2\right], \nonumber\\
A' \!\!\!\!&=&\!\!\!\! - A + \frac{\alpha'}{2} \D^2 \Phi\,,
\end{eqnarray}
to get (omitting the primes for $A'$ and $B'$) 
\begin{equation}\label{cosmoHSZ_action_simplest}
\begin{split}
I = \int dt \, n \, e^{-\Phi} \, \Big\{ &\frac{1}{\alpha'}\Big[A B + \frac12 \Tr{\Z} -\frac16 \Tr{\Z^3}\Big] - \frac18 \Tr{(\D \Z)^2}  - (\D \Phi)^2
 \\
&+ \frac{\alpha'{}^2}{4}\Big[\frac14 (\D^3\Phi)^2 + \frac13 (\D^2\Phi)^3\Big]\Big\}\;. 
\end{split}
\end{equation}

We observe that after the above series of field redefinitions  $A$ and $B$ completely trivialize in the sense that their equations of motion simply set them to zero. Therefore, the original theory given by \eqref{cosmoHSZ_lagrangian2} is equivalent to an effective theory for $\Z, n$ and $\Phi$ only, whose action is given by \eqref{cosmoHSZ_action_simplest} after setting $A=B=0$. As a consequence, from now on we can use this effective action to explore the physics of the $\Z,n,\Phi$ system alone, which in the following section we
take as a starting point to get an effective action for standard gravity fields. In section \ref{sec_2dev}, however, we will reintroduce $A$ and $B$ to see that, even though they encode no extra information for the physical fields, they can still be used to bring the theory to a formulation without positive powers of $\alpha'$, i.e., without more than two derivatives.

\section{Higher-derivative corrections at order $\alpha'^{\,4}$}

Our goal in this section is to relate the cosmological reduction of HSZ theory to a conventional gravity 
theory  featuring only the metric, B-field and dilaton. 
To this end, setting $d = D - 1$, we express the internal double metric $\Z$ in terms of an $O(d, d)$ valued 
generalized metric, encoding the metric and the $B$-field, plus extra fields. 
Integrating out these extra fields then induces higher-derivative $\alpha'$ corrections for the remaining fields that we compute to order $\alpha'^4$. We end the section by  showing how to bring the resulting action to the minimal form classified by Hohm and Zwiebach \cite{Hohm:2015doa}.

\subsection{From the double to the generalized metric}\label{ZtoS}

We start from the action in the form \eqref{cosmoHSZ_action_simplest}, setting  $A=B=0$, as implied by their own field equations. 
Thus, we consider the following action for the remaining fields $\Z, n$ and $\Phi$: 
\begin{equation}\label{cosmoHSZ_effective}
\begin{aligned}
I &= \int dt \, n \, e^{-\Phi} \, \left\{ \frac{1}{2 \alpha'} \left[ \Tr{\Z} -\frac13 \Tr{\Z^3}\right] - \frac18 \Tr{(\D \Z)^2} - (\D \Phi)^2 + \frac{\alpha'{}^2}{4} \left[ \frac14 (\D^3 \Phi)^2 + \frac13 (\D^2 \Phi)^3\right] \right\} \,. 
\end{aligned}
\end{equation}
From the  general  variation of this action
\begin{equation}\label{GeneralVAR}
\delta I = \int dt \, n \, e^{-\Phi} \left[\Tr{\delta \Z E_{\Z}} + \delta \Phi E_{\Phi} + \frac{\delta n}{n} E_{n} \right]\,, 
\end{equation}
one obtains the equations of motion
\begin{subequations}\label{EOMZsystem}
\begin{align}
0 = E_{\Z} &\equiv \frac{1}{2 \alpha'}\left( 1 - \Z^2\right) - \frac14 \D\Phi\D \Z + \frac14\D^2\Z\label{EOMZ}\,, \\
0 = E_{\Phi} &\equiv -\frac{1}{2 \alpha'}\left[\Tr{\Z} - \frac13 \Tr{\Z^3}\right] + \frac18 \Tr{(\D \Z)^2} - (\D \Phi)^2 +2 \D^2 \Phi \nn \\
&\quad -\frac{\alpha'{}^2}{8} \left[ \vphantom{\frac 1 2}\D^6 \Phi - 3 \D\Phi\D^5\Phi -7\D^2\Phi\D^4\Phi +3(\D\Phi)^2\D^4\Phi\right.-\frac92(\D^3\Phi)^2 +11\D\Phi\D^2\Phi\D^3\Phi  \nn\\ 
& \quad \quad \quad \quad - (D\Phi)^3\D^3\Phi  \left.\quad +\frac83(\D^2\Phi)^3 -2(\D\Phi)^2(\D^2\Phi)^2\right] \label{EOMphi}\,, \\
0 = E_{n} &\equiv \frac{1}{2 \alpha'}\left[\Tr{\Z} - \frac13 \Tr{\Z^3}\right] + \frac18 \Tr{(\D \Z)^2} + (\D \Phi)^2 \nn \\
&\quad -\frac{\alpha'{}^2}{8} \left[ \vphantom{\frac 1 2}\D\Phi\D^5\Phi -\D^2\Phi\D^4\Phi -2(\D\Phi)^2\D^4\Phi\right. +\frac12(\D^3\Phi)^2 -4\D\Phi\D^2\Phi\D^3\Phi \nn\\ 
& \quad  \quad \quad \quad+ (D\Phi)^3\D^3\Phi  \left. +\frac43(\D^2\Phi)^3 +2(\D\Phi)^2(\D^2\Phi)^2\right]\,,  \label{EOMn}
\end{align}
\end{subequations}
where for the first equation we used a matrix notation for $\Z_\bullet{}^\bullet$. As a consistency check, we verified  that 
these equations  satisfy the Bianchi/Noether  identity implied by gauge invariance under  time reparameterizations. 
To this end one specializes in (\ref{GeneralVAR})  the variations to the one-dimensional diffeomorphism transformations
\begin{equation}
\delta \Phi = \xi \dot{\Phi} \,, \quad  \quad \delta n = \partial_{t}(\xi n)\,,  \quad  \quad \delta \Z = \xi \dot{\Z}\,, 
\end{equation}
which yields
\begin{equation}
\delta_\xi I = \int dt \, n \, e^{-\Phi} \xi \left[ \Tr{\dot{\Z} E_{\Z}} + \dot{\Phi} E_{\Phi} - e^{\Phi} \partial_t 
\left( e^{-\Phi} E_n\right)\right] = 0\,.
\end{equation}
Gauge invariance implies that this must hold for arbitrary $\xi$, from which we infer: 
\begin{equation}\label{BI}
\D \Phi (E_\Phi + E_n) + \Tr{\D \Z E_{\Z}} - \D E_n = 0\,. 
\end{equation}
This  is indeed identically satisfied by \eqref{EOMZsystem}.

Even though discussing solutions of \eqref{EOMZsystem} is beyond the scope of this work, there are two of them that can be easily found. The first one is the trivial configuration
\begin{equation}
\Z(t)_M{}^N = (S_0)_M{}^N\,, \quad \Phi(t) = \Phi_0\,,
\end{equation}
where $S_0$ is the standard generalized metric but with constant entries, and $\Phi_0$ is also a constant. In that case \eqref{EOMZsystem} is trivially satisfied. More surprisingly, however, is the non-trivial and \textit{non-perturbative} solution
\begin{equation}\label{solutionZ}
\Z(t)_M{}^N = - \delta_M{}^N\,, \quad \Phi(t) = \sqrt{\frac{2 d}{3 \alpha'}}\, t + \Phi_0\,.
\end{equation}
Note that  $\Z(t)_M{}^N = \delta_M{}^N$ is not a solution for a real dilaton field.
We consider this solution to belong to a separate and somewhat pathological branch 
that is incompatible with the interpretation as a generalized metric featuring a conventional spacetime metric plus 
B-field. For this reason we will focus in later sections on solutions with $\Z\neq -{\bf 1}$. 

We now turn to the decomposition of the double metric into a generalized metric that encodes conventional 
gravity fields plus extra fields that will be integrated out perturbatively\cite{Hohm:2015mka}. 
We write 
\begin{equation}\label{ZSF}
\Z = S + F\,, 
\end{equation}
where $S \in O(d, d)$ is the internal generalized scalar metric satisfying $S^2 = 1$. 
A generalized metric can be parametrized in terms of  symmetric and antisymmetric tensors $g$ and $b$, respectively, as 
follows 
\begin{equation}
	S \equiv  \begin{pmatrix}
	b g^{-1} & g - b g^{-1} b\\
	g^{-1} & - g^{-1} b
	\end{pmatrix}  .
	\end{equation}
The matrix $F$ encodes the  extra fields and can be 
assumed to obey some constraints, which we now describe. We first recall that with the 
generalized metric one can build projectors
\begin{equation}
P \equiv \frac12\left(1 - S\right) \,, \qq \bar P \equiv \frac12\left(1 + S\right)\,, 
\end{equation}
satisfying 
 \begin{equation}
 P^2 = P\,, \qq \bar P^2 = \bar P\,,  \qq P \bar P = \bar P P = 0\,. 
 \end{equation} 
Furthermore, we have the following useful identities 
\begin{equation}
P S = S P = - P \,, \qq \bar P S = S \bar P = \bar P\,. 
\end{equation}
A generic matrix $A = A_\bullet{}^\bullet$ then can be projected into $\pm$ components, defined as
\begin{equation}
A_+ \equiv P A P + \bar P A \bar P\,,  \qq A_{-} \equiv P A \bar P + \bar P A P\,, 
\end{equation}
{such that} 
\begin{equation} 
 A = A_+ + A_-\,. 
\end{equation} 
We also have the relations 
\begin{equation}\label{+-projections}
A_{\pm} = \frac12\left(A \pm S A S\right)\,, \qq A_{\pm} S = \pm S A_{\pm} \,. 
\end{equation}
Such relations are useful in order to show that certain traces vanish: 
\begin{equation}\label{Trace-}
\Tr{A_-} = \Tr{A_- S^2} = - \Tr{S A_- S} = - \Tr{A_- S^2} = - \Tr{A_-} = 0\,, 
\end{equation}
where we used $S^2 = 1$, \eqref{+-projections} and the cyclicity of the trace. Thus, traces of minus-projected tensors vanish. 

While the decomposition \eqref{ZSF} is totally generic, in perturbation theory we think of $F$ as being of one order in $\alpha'$ higher than $S$. This is motivated by the fact that, in this case, the leading order contribution to \eqref{cosmoHSZ_effective} does not depend on $F$ and it corresponds to the standard Neveu-Schwarz sector of supergravity in cosmological backgrounds. We can then assume $F$ to be a constrained field belonging to the $+$ subspace, i.e.~$F = F_+$, because any part in $F$ belonging to the $-$ subspace can be removed by a field redefinition 
$S \rightarrow S + \delta S$ since $ \delta S = [\delta S]_-$  
as follows by taking the variation of $S^2=1$. 
Thus, without loss of generality, in perturbation theory we can write 
\begin{equation}\label{finalZsplit}
\Z = S + F_+\;. 
\end{equation}

Let us now inspect  the equation of motion \eqref{EOMZ} after using this decomposition of $\Z$. 
Inserting  (\ref{finalZsplit}) into  \eqref{EOMZ} one obtains 
\begin{equation}\label{EOMZsplit}
S F_+ = \frac{\alpha'}{4} \Box_\Phi(S + F_+) - \frac12 F_+^2\,, 
\end{equation}
where here and in the following it is convenient to introduce  the linear operator
\begin{equation}\label{Boxphi}
\Box_\Phi \equiv \D^2 - \D \Phi \D\,. 
\end{equation}
From this we can obtain  the equations for $F_+$ or $S$ by projecting  into $\pm$ subspaces. 
For the generalized metric we project to the $-$ subspace, and use $[F_+]_- = [F_+^2]_- = 0$. This yields 
\begin{equation}
\left[ \Box_\Phi(S + F_+)\right]_{-} = 0\;. 
\end{equation}
For the extra fields we take the $+$ projection: 
\begin{equation}\label{EOMF}
F_+ = \frac{\alpha'}{4}[\Box_\Phi S]_+ S + \frac{\alpha'}{4}[\Box_\Phi F_+]_+ S - \frac12 F_+^2 S\;. 
\end{equation}
Since this is the equation of motion for $F_+$, we can solve for it  perturbatively in $\alpha'$ following the iterative procedure described in the following section.

\subsection{Integrating out the extra fields}

Equation \eqref{EOMF} is the starting point for integrating out the extra fields $F_+$. We 
assume a perturbative expansion in $\alpha'$, namely
\begin{equation}\label{alphaprimeFexp}
F_+ = \sum_{i \ge 1} \alpha'{}^i F_+^{(i)}\;. 
\end{equation} 
Inserting  this expansion into \eqref{EOMF} we obtain the recursive relations
\begin{equation}\label{F+}
\begin{aligned}
F_+^{(1)} &= \frac14 [\Box_\Phi S]_+ S = - \frac14 (\D S)^2\,, \\
F_+^{(i)} &= \frac14 [\Box_\Phi F_+^{(i-1)}]_+ S - \frac12 \sum_{j = 1}^{i - 1} F_+^{(j)} F_+^{(i - j)} S\,,  \qq i \geq 2\,. 
\end{aligned}
\end{equation}
By solving these equations recursively, we can express all $F_+^{(i)}$ in terms of $S$, $\Phi$ and $n$. Plugging these expressions  
back into the action \eqref{cosmoHSZ_effective} yields the effective action for the conventional fields.

We perform this computation explicitly to  order $\alpha'^4$ by first inserting  the decomposition $\Z = S + F_+$ into the action (\ref{cosmoHSZ_effective}), which then splits as
\begin{equation}\label{generalAlphaPrimeExp}
I = I^{(0)} + \alpha' I^{(1)} + \alpha'{}^2 I^{(2)} + \alpha'{}^3 I^{(3)} + \alpha'{}^4 I^{(4)} + \mathcal{O}(\alpha'{}^5)\,, 
\end{equation}
with
\begin{equation}\label{Salpha4}
\begin{aligned}
I^{(0)} &= \int dt \, n \, e^{-\Phi} \, \left\{ - \frac18 \Tr{(\D S)^2} - (\D \Phi)^2 \right\}\,, \\
I^{(1)} &= \int dt \, n \, e^{-\Phi} \, \left\{ - \frac12 \Tr{S (F_+^{(1)})^2} -\frac14 \Tr{\D S \D F_+^{(1)}}\right\}\,, \\
I^{(2)} &= \int dt \, n \, e^{-\Phi} \, \left\{- \frac16 \Tr{(F_+^{(1)})^3} - \frac18 \Tr{(\D F_+^{(1)})^2} + \frac{1}{16} (\D^3 \Phi)^2 + \frac{1}{12} (\D^2 \Phi)^3\right.\\
&\left. \hspace{2.4cm} - \Tr{S F_+^{(1)} F_+^{(2)}}  - \frac14 \Tr{\D S \D F_+^{(2)}}  \right\}\,, \\
I^{(3)} &= \int dt \, n \, e^{-\Phi} \, \left\{- \frac12 \Tr{S (F_+^{(2)})^2} - \frac12 \Tr{F^{(2)}_+ (F_+^{(1)})^2} -\frac14 \Tr{\D F_+^{(1)} \D F_+^{(2)}}\right.\\
&\left. \hspace{2.4cm} - \Tr{S F_+^{(1)} F_+^{(3)}}  - \frac14 \Tr{\D S \D F_+^{(3)}}  \right\}\,, \\
I^{(4)} &= \int dt \, n \, e^{-\Phi} \, \left\{- \frac12 \Tr{F_+^{(1)} (F_+^{(2)})^2} - \frac18 \Tr{(\D F_+^{(2)})^2} \right.\\
& \hspace{2.4cm}- \Tr{S F_+^{(1)} F_+^{(4)}} - \frac14 \Tr{\D S \D F_+^{(4)}}\\ 
&\left. \hspace{2.4cm} - \Tr{S F_+^{(2)} F_+^{(3)}} - \frac12 \Tr{(F_+^{(1)})^2 F_+^{(3)}} - \frac14 \Tr{\D F_+^{(1)} \D F_+^{(3)}} \right\}\,. 
\end{aligned}
\end{equation}

With these expressions, naively it seems that we need to solve the recursive equation up to quartic order to get the final expression of the action in terms of $S$. However, now we proceed to show that \eqref{Salpha4} can be simplified and in particular we can remove any appearance of $F_+^{(3)}$ and $F_+^{(4)}$. We describe in detail how to simplify $I^{(1)}$, since the exact same simplifications can be performed in the higher orders. We find
\begin{equation}
\begin{aligned}
I^{(1)} &= \int dt \, n \, e^{-\Phi} \, \left\{ - \frac12 \Tr{S (F_+^{(1)})^2} -\frac14 \Tr{\D S \D F_+^{(1)}}\right\} \\
&= \int dt \, n \, e^{-\Phi} \, \left\{ - \frac12 \Tr{S (F_+^{(1)})^2} +\frac14 \Tr{\Box_\Phi S F_+^{(1)}}\right\}\\
&= \int dt \, n \, e^{-\Phi} \, \left\{ - \frac12 \Tr{S (F_+^{(1)})^2} +\frac14 \Tr{[\Box_\Phi S]_+ F_+^{(1)}}\right\}\\
&= \int dt \, n \, e^{-\Phi} \, \Tr{\left[-\frac12 S F_+^{(1)} + \frac14 [\Box_\Phi S]_+\right] F_+^{(1)}} \\
&= \frac12 \int dt \, n \, e^{-\Phi} \, \Tr{S (F_+^{(1)})^2} \,,
\end{aligned}
\end{equation}
where in the second equality we integrated by parts and used the definition \eqref{Boxphi}. From the second to third line we use 
that only the $+$ projection survives inside the trace due to \eqref{Trace-}. For the final step one uses 
the leading solution \eqref{F+} for $F_+^{(1)}$. Following these steps in the higher corrections leads to the following simplifications
\begin{equation}\label{Salpha4final}
\begin{aligned}
I^{(0)} &= \int dt \, n \, e^{-\Phi} \, \left\{ - \frac18 \Tr{(\D S)^2} - (\D \Phi)^2 \right\}\,, \\
I^{(1)} &=\frac12 \int dt \, n \, e^{-\Phi} \, \Tr{S (F_+^{(1)})^2}\,,\\
I^{(2)} &= \int dt \, n \, e^{-\Phi} \, \left\{- \frac16 \Tr{(F_+^{(1)})^3} - \frac18 \Tr{(\D F_+^{(1)})^2} + \frac{1}{16} (\D^3 \Phi)^2 + \frac{1}{12} (\D^2 \Phi)^3\right\}\,,\\
I^{(3)} &=\frac12 \int dt \, n \, e^{-\Phi} \, \Tr{S (F_+^{(2)})^2}\,,\\
I^{(4)} &= \int dt \, n \, e^{-\Phi} \, \left\{- \frac12 \Tr{F_+^{(1)} (F_+^{(2)})^2} - \frac18 \Tr{(\D F_+^{(2)})^2} \right\}\,. 
\end{aligned}
\end{equation}
This rewriting tells us that only $F_+^{(1)}$ and $F_+^{(2)}$ are needed,  which are given by \eqref{F+} for $i=1,2$, 
\begin{equation}
\begin{aligned}
F_+^{(1)} &= - \frac14 (\D S)^2\,, \\
F_+^{(2)} &=  -\frac{1}{16} [\Box_\Phi(\D S \D S)]_+ S - \frac{1}{32} (\D S)^4 S\,. 
\end{aligned}
\end{equation}
 
In order to evaluate the action \eqref{Salpha4final} for these solutions we found it convenient to  write the result  in terms of 
\begin{equation}
\L \equiv \D S S\,, 
\end{equation}
which satisfies
\begin{equation}\label{L-}
\L = [\L]_-\ , \ \  \qq \D \L = [\D \L]_-\,. 
\end{equation}
In particular, we will see that the zeroth order equations  for $S$ take a simpler form when written in terms of $\L$. The lowest order solutions for the extra fields then read
\begin{equation}
\begin{aligned}
F^{(1)}_+ &= \frac14 \L^2\,, \\
F^{(2)}_+ &= - \frac{1}{32} \L^4 S + \frac{1}{16} [\Box_\Phi(\L^2)]_+ S\,. 
\end{aligned}
\end{equation}
Inserting these expressions into \eqref{Salpha4final} one obtains an action of the form (\ref{generalAlphaPrimeExp}) 
with 
\begin{equation}\label{Salpha4finalL}
\begin{aligned}
I^{(0)} &= \int dt \, n \, e^{-\Phi} \, \left\{ \frac18 \Tr{\L^2} - (\D \Phi)^2 \right\}\,, \\
I^{(1)} &= 0 \,, \\
I^{(2)} &= \int dt \, n \, e^{-\Phi} \, \left\{- \frac{1}{3.2^7} \Tr{\L^6} - \frac{1}{2^7} \Tr{\D (\L^2) \D (\L^2)} + \frac{1}{16} (\D^3 \Phi)^2 + \frac{1}{12} (\D^2 \Phi)^3\right\}\,, \\
I^{(3)} &= \int dt \, n \, e^{-\Phi} \, \Ld_3(\D \L, \D \Phi, \Tr{\L^2})\,,\\
I^{(4)} &= \int dt \, n \, e^{-\Phi} \, \Ld_4(\D \L, \D \Phi, \Tr{\L^2})\,,
\end{aligned} 
\end{equation}
where we used
\begin{equation}
\Tr{S \L^n} = 0\,, \quad n \in \mathbb{N}_0\,,
\end{equation}
which follows  by anticommuting $S$ with  $\L$ and using the cyclicity of the trace.
The notation  $\Ld_i(\D \L, \D \Phi, \Tr{\L^2})$ refers to functions of the given arguments, whose specific form is irrelevant for our purposes. As discussed in previous papers \cite{Hohm:2019jgu, Codina:2021cxh}, terms containing derivatives of $\L$, 
dilaton terms or $\Tr{\L^2}$ can be removed by field redefinitions at the expense of introducing higher order terms. We then call such terms ``removable''. The process of implementing field redefinitions is cumbersome and has been discussed previously \cite{Codina:2021cxh}, so we move the general procedure to the Appendix. Here we just summarize our findings. A redefinition of order $4$ eliminates $I^{(4)}$ because it is purely removable. Of course it might induce higher order terms, that we are neglecting here. A redefinition of order $3$ eliminates $I^{(3)}$ for the same reason. This could reintroduce new terms in $I^{(4)}$, but it does not because there is no $I^{(1)}$ and so the first order correction to the equations of motion vanishes. So we can then start by eliminating $I^{(3)}$ and $I^{(4)}$ completely to order $\alpha'^4$. When it comes to removing the last three terms in $I^{(2)}$, the second order equations of motion reintroduce terms in $I^{(4)}$, some which are removable at no cost, plus an extra term that defines the new coefficients found in this paper:  $c_{5,0}$ and $c_{5,1}$. The final result of this systematic procedure is the HSZ action in the cosmological classification
\begin{equation}\label{HSZcosmoclassification}
I = \int dt \, n \, e^{-\Phi} \, \left\{ \frac18 \Tr{\L^2} - (\D \Phi)^2 - \alpha'{}^2 \frac{1}{3 \cdot  2^7} \Tr{\L^6} + \alpha'{}^4 \frac{1}{3\cdot 2^{13}} \Tr{\L^4}\Tr{\L^6}  + \mathcal{O}(\alpha'{}^5)\right\}\,.
\end{equation}
By coming back to the original basis in terms of $\D S$ and the classification given in \cite{Hohm:2019jgu}, 
according to which the action must take the form 
\be\label{MinimalActionCoeff}
 \begin{split}
I  = \int dt \, e^{- \Phi} \Big[&- \dot \Phi^2 + c_{1,0} \, \Tr{(\D S)^2} 
+ \alpha'\, c_{2,0}\, \Tr{(\D S)^4} + \alpha'{}^2 \,c_{3,0}\,  \Tr{(\D S)^6} \\
&+ \alpha'{}^3 \left(c_{4,0}\, \Tr{(\D S)^8} + c_{4,1} \,(\Tr{(\D S)^4})^2\right) \\
&+ \alpha'{}^4 \left(c_{5,0}\, \Tr{(\D S)^{10}} + c_{5,1} \,\Tr{(\D S)^6}\, \Tr{(\D S)^4}\right) \vphantom{\frac 1 8}\Big] \ , 
 \end{split}
 \ee
we obtain by comparison of coefficients
\begin{center}
	\begin{tabular}{|| c || c || c || c || c || c || c ||}
		\hline
		$c_{1,0}$ & $c_{2,0}$ & $c_{3,0}$ & $c_{4, 0}$ & $c_{4, 1}$ & $c_{5, 0}$ & $c_{5, 1}$\\ [0.5ex]
		\hline
		$-\frac18$ & $0$ & $\frac 1 {3 \cdot 2^7}$  & $0$ & $0$ & $0$ & $-\frac 1 {3 \cdot 2^{13}}$\\
		\hline
	\end{tabular}
\end{center}
Up to and including $c_4$ the coefficients coincide with the ones obtained in \cite{Codina:2021cxh}, whereas $c_{5, 0}$ and $c_{5, 1}$ are new results.

\section{Two-derivative reformulation}\label{sec_2dev}

In \cite{Hohm:2016lim}, HSZ theory was considered up to quadratic order in field perturbations about flat space. 
In this limit the higher-derivative terms can be removed  by introducing certain auxiliary fields. 
One can then take the tensionless limit $\alpha' \rightarrow \infty$ smoothly, for  which one finds an enhanced gauge symmetry whose corresponding  gauge fields are the auxiliary fields. The objective of this section is two-fold: First, we will show that in the cosmological setting the auxiliary fields introduced in \cite{Hohm:2016lim} are not required in order to remove higher orders in $\alpha'$, since the already present $A$ and $B$ can be used to accomplish this. Second, for completeness,   we show that the auxiliary fields are  nevertheless needed  in order to make contact with the formulation of \cite{Hohm:2016lim}, but we will see 
that for the cosmological setting the enhanced gauge symmetries can be trivialized in the sense that the fields  
can be rearranged  into gauge invariant variables.
 
\subsection{Two-derivative theory}

The action \eqref{cosmoHSZ_action_simplest} contains only two terms at higher orders in $\alpha'$ given by the terms 
in the second line. Remarkably, both terms can be removed by exact field redefinitions of $A$ and $B$. Explicitly, by performing the transformations
\begin{equation}
\begin{aligned}
B &= B' - \frac23 A'{}^2 + \alpha'\left[\frac14 \D A' \D \Phi - \frac14 \D^2 A' + \frac13 A' \D^2 \Phi\right]\\
& \hspace{2.2cm} - \frac{\alpha'{}^2}{2}\left[-\frac14 \D^4 \Phi + \frac14 \D\Phi \D^3 \Phi + \frac13 (\D^2\Phi)^2\right]\,, \\
A &= A' + \frac{\alpha'}{2} \D^2 \Phi \,, 
\end{aligned}
\end{equation}
the action \eqref{cosmoHSZ_action_simplest} can be brought to the following  two-derivative form
(omitting the primes) 
\begin{equation}\label{cosmoHSZ_action_2derivative}
\begin{split}
I = \int dt \, n \, e^{-\Phi} \, \Big\{ &\frac{1}{\alpha'}\Big[A B - \frac23 A^3 + \frac12 \Tr{\Z} -\frac16 \Tr{\Z^3}\Big] \\
 &- \frac18 \Tr{(\D \Z)^2}  - (\D \Phi)^2 + \frac14 (\D A)^2 + \frac12 B \D^2 \Phi\Big\}\,. 
 \end{split}
\end{equation}
In this formulation, all higher orders in $\alpha'$ are `hidden' in the on-shell value of $A$ and $B$. Indeed, we note  that $B$ enters in the action just linearly, so it plays the role of a Lagrange multiplier, which imposes a condition on the scalar field $A$: \begin{equation}\label{EOMB}
\frac{\delta I}{\delta B} = 0 \quad \Rightarrow \quad A = -\frac{\alpha'}{2} \D^2 \Phi\,. 
\end{equation}
Reinserting this value of $A$ into the action \eqref{cosmoHSZ_action_2derivative}, the $A^3$ and $(\D A)^2$ couplings are replaced by  the $\alpha'{}^2$ terms $(\D^2 \Phi)^3$ and $(\D^3 \Phi)^2$, respectively. The resulting action is exactly the effective action for $\Z, n$ and $\Phi$ obtained from \eqref{cosmoHSZ_action_simplest} after fixing $A=B=0$.

In order to relate to the results in  \cite{Hohm:2016lim} it is instructive to go to a more democratic formulation in terms of two `massive' scalar fields $A^+$ and $A^-$. To do so, we first integrate by parts the term $B \D^2 \Phi$ and perform the following 
field redefinitions: 
\begin{equation}
\begin{aligned}
A &= -(A^+ + A^-)\,,\qquad 
B = 2(A^+ - A^-)\,, 
\end{aligned}
\end{equation}
to arrive at
\begin{equation}\label{cosmoHSZ_action_massiveApm1}
\begin{split}
I = \int dt \, n \, e^{-\Phi} \, \Big\{ &\frac{1}{\alpha'}\Big[ - 2 (A^+)^2 + 2 (A^-)^2 + \frac23 (A^+ + A^-)^3 + \frac12 \Tr{\Z} -\frac16 \Tr{\Z^3}\Big] \\
&  - \frac18 \Tr{(\D \Z)^2}  - (\D \Phi)^2  + \frac14 (\D A^+)^2 + \frac14 (\D A^-)^2  
\\
& + \frac12 \D A^+ \D A^- - \D(A^+ - A^-) \D \Phi + (A^+ - A^-)(\D\Phi)^2 \Big\}\,. 
\end{split}
\end{equation}
Finally we remove the mixed terms with a field redefinition of the dilaton, 
\begin{equation}
\Phi = \Phi' - \frac12(A^+ - A^-)\,. 
\end{equation} 
Omitting the primes for the transformed dilaton, the final action reads
\begin{equation}\label{cosmoHSZ_action_massiveApm2}
\begin{split}
I = \int dt \, n \, e^{-\Phi + \frac12(A^+ - A^-)} \, &\left\{ \frac{1}{\alpha'}\left[ \frac23 (A^+ + A^-)^3 + \frac12 \Tr{\Z} -\frac16 \Tr{\Z^3}\right]\right.\\
&- \frac18 \Tr{(\D \Z)^2}  - (\D \Phi)^2 \\
& + \frac12 (\D A^+)^2 - \frac{2}{\alpha'} (A^+)^2 + \frac12 (\D A^-)^2  + \frac{2}{\alpha'} (A^-)^2\\
& \left.  + (A^+ - A^-)\left[\D(\Phi - \frac12(A^+ - A^-))\right]^2\right\}\,.
\end{split}
\end{equation}
Here we recognize from the quadratic terms in the third line  that $A^{\pm}$ correspond to two `massive' scalar fields with 
masses $ (M^{\pm})^2 = \mp \frac{4}{\alpha'}$, respectively.

Once we have the action in a form with no positive $\alpha'$ powers we can take  the tensionless limit 
$\alpha' \rightarrow \infty$. Then all $\frac{1}{\alpha'}$ terms drop from the action, in particular the mass terms for $A^{\pm}$. The resulting action is given by
\begin{equation}\label{cosmoHSZ_action_infty1}
\begin{aligned}
I_{\infty} = \int dt \, n \, e^{-\Phi + \frac12(A^+ - A^-)} \, &\left\{ - \frac18 \Tr{(\D \Z)^2}  - (\D \Phi)^2 + \frac12 (\D A^+)^2 + \frac12 (\D A^-)^2 \right.\\
& \left. + \, (A^+ - A^-)\left[\D(\Phi - \frac12(A^+ - A^-))\right]^2\right\}\,. 
\end{aligned}
\end{equation}
In contrast to the enhanced gauge symmetries uncovered in \cite{Hohm:2016lim}, here the tensionless limit does not appear to exhibit any additional gauge symmetries. However, it does have a symmetry in the form of a constant shift $A^{\pm} \rightarrow a = $ const. because, apart from the kinetic term, $A^{\pm}$ appear in the combination $A^+ - A^-$, in contrast to the full theory  \eqref{cosmoHSZ_action_massiveApm2} with finite  $\alpha'$.  We will show in the next section that upon 
introducing two extra auxiliary fields,  motivated by \cite{Hohm:2016lim}, we can display the expected enhanced gauge symmetries, 
albeit in a somewhat trivial fashion.

\subsection{Tensionless limit}

In the previous section we have seen that introducing additional auxiliary fields is not required in order to remove higher orders in $\alpha'$. Nevertheless, here we will introduce two extra fields  with the  goal of identifying  the enhanced symmetry  found in \cite{Hohm:2016lim}, after taking the tensionless limit.

Returning to \eqref{cosmoHSZ_action_simplest} let us  `integrate-in' two auxiliary fields $\varphi$ and $\bar{\varphi}$ in order to remove the terms of order  $\alpha'{}^2$, mimicking the procedure  in \cite{Hohm:2016lim}: 
\begin{equation}\label{cosmoHSZ_action_varphis}
\begin{aligned}
\int dt \, n \, e^{-\Phi} \, &\left\{ \frac{1}{\alpha'}\left[A B - 2 \varphi^2 + 2 \bar{\varphi}^2 + \frac23 (\varphi + \bar{\varphi})^3 + \frac12 \Tr{\Z} -\frac16 \Tr{\Z^3} \right]\right.\\
& \left. - \frac18 \Tr{(\D \Z)^2}  - (\D \Phi)^2 + \frac14 (\D \varphi)^2 + \frac14 (\D \bar\varphi)^2 + \frac12 \D \varphi \D \bar \varphi \right.\\ 
&\left.- \D(\varphi - \bar \varphi)\D \Phi +  (\varphi - \bar \varphi) (\D \Phi)^2\right\}\,. 
\end{aligned}
\end{equation}
In order to show that this action is equivalent to \eqref{cosmoHSZ_action_simplest} we  compute the equations of motion 
for $\varphi$ and $\bar \varphi$, which yields 
\begin{equation}
\begin{aligned}
\varphi &= \frac{\alpha'}{4} \D^2 \Phi + \eta\,, \qquad 
\bar \varphi = \frac{\alpha'}{4} \D^2 \Phi - \eta\,,\\
\eta &\equiv \frac12 (\varphi + \bar \varphi)^2 + \frac{\alpha'}{8}\left[ \D(\varphi + \bar \varphi) \D \Phi 
- \D^2(\varphi + \bar \varphi)\right]\,.
\end{aligned}
\end{equation}
Upon reinsertion into \eqref{cosmoHSZ_action_varphis},  all $\eta$ contributions cancel, and the only surviving terms are exactly the two dilaton contributions at order $\alpha'{}^2$ of \eqref{cosmoHSZ_action_simplest}. Another simpler way to see that this theory is equivalent to the original one is by noticing that \eqref{cosmoHSZ_action_varphis} is equal to  \eqref{cosmoHSZ_action_massiveApm1} after identifying $\varphi \rightarrow A^+$ and $\bar \varphi \rightarrow A^-$ and setting $A=B=0$.
Moreover, as we did in the previous section, we can perform a dilaton field redefinition 
\begin{equation}
\Phi = \Phi' - \frac12(\varphi - \bar \varphi)\,, 
\end{equation}
to arrive at the analogue  of \eqref{cosmoHSZ_action_massiveApm2}, namely
\begin{equation}\label{cosmoHSZ_action_varphis2}
\begin{aligned}
I = \int dt \, n \, e^{-\Phi + \frac12(\varphi - \bar \varphi)} \, &\left\{ \frac{1}{\alpha'}\left[ A B +  \frac23 (\varphi + \bar \varphi)^3 + \frac12 \Tr{\Z} -\frac16 \Tr{\Z^3}\right]\right.\\
&- \frac18 \Tr{(\D \Z)^2}  - (\D \Phi)^2\\
&+ \frac12 (\D \varphi)^2 - \frac{2}{\alpha'} \varphi^2 + \frac12 (\D \bar \varphi)^2  + \frac{2}{\alpha'} \bar{\varphi}^2\\
&\left. + (\varphi - \bar \varphi)\left[\D(\Phi - \frac12(\varphi - \bar \varphi))\right]^2\right\}\,. 
\end{aligned}
\end{equation}
In contrast to \eqref{cosmoHSZ_action_massiveApm2}, here we still have $A$ and $B$, and 
we can perform the following field redefinitions\footnote{Here $\bar A$ plays the role of a $B'$.}
\begin{equation}
\begin{aligned}
A &= A' + \bar A - (\varphi' - \bar \varphi')\,, \quad &&B = A' - \bar A - (\varphi' + \bar \varphi')\,, \\
\varphi &= \varphi' - \frac12 A'\,, \quad &&\bar \varphi = \bar \varphi' 
+ \frac12 \bar A\,, \\
\end{aligned}
\end{equation}
to arrive at the action
\begin{equation}\label{cosmoHSZ_action_varphis3}
\begin{aligned}
I = \int dt \, n \, e^{-\Phi + \frac12(\varphi - \bar \varphi) - \frac14(A + \bar A)} \, &\left\{ \frac{1}{\alpha'}\left[  \frac23 (\varphi + \bar \varphi -\frac12(A - \bar A))^3 + \frac12 \Tr{\Z} -\frac16 \Tr{\Z^3}\right]\right.\\
& - \frac18 \Tr{(\D \Z)^2}  - (\D \Phi)^2\\
& + \frac18 (\D A)^2 + \frac12 (\D \varphi)^2 - \frac12 \D A \D \varphi + \frac{1}{\alpha'}\left[\frac12 A^2 - \varphi^2\right]\\ 
& + \frac18 (\D \bar A)^2 + \frac12 (\D \bar \varphi)^2 - \frac12 \D \bar A \D \bar \varphi - \frac{1}{\alpha'}\left[\frac12 \bar A^2 - \bar \varphi^2\right]\\
& \left. + \left[\varphi - \bar \varphi -\frac12(A + \bar A)\right]\left[\D(\Phi - \frac12(\varphi - \bar \varphi) + \frac14 (A + \bar A))\right]^2\right\}\,, 
\end{aligned}
\end{equation}
omitting again the primes of the transformed fields. One may verify  that the quadratic approximation for  field perturbations around the classical solution $A = \bar A = \varphi = \bar \varphi = 0$ coincides with the one obtained in equation (5.1) of \cite{Hohm:2016lim}, once reduced to a cosmological background, where  the components $a_{0 0}, \bar a_{0 0}, \varphi_{0 0}$ and $ \bar \varphi_{0 0}$ correspond to the fluctuations of $A, \bar A, \varphi$ and $\bar \varphi$, respectively.

Given  this match, we expect to recover the enhanced gauge symmetry in the tensionless limit. To this end consider the following  field redefinition of the lapse function: 
\begin{equation}
n = n' e^{-\Phi + \frac12(\varphi - \bar \varphi) - \frac14(A + \bar A)}\,, 
\end{equation}
which removes completely the exponential factor for the two-derivative part of the action, at the expense 
of modifying the measure of the $\frac{1}{\alpha'}$ terms. But the latter terms   disappear 
in the limit $\alpha' \rightarrow \infty$, and 
we obtain 
\begin{equation}\label{cosmoHSZ_action_infty2}
\begin{aligned}
I_{\infty} = \int dt \, n \, &\left\{ - \frac18 \Tr{(\D \Z)^2}  - (\D \Phi)^2 \right.\\
& + \frac18 (\D A)^2 + \frac12 (\D \varphi)^2 - \frac12 \D A \D \varphi + \frac18 (\D \bar A)^2 + \frac12 (\D \bar \varphi)^2 - \frac12 \D \bar A \D \bar \varphi \\
& \left. + \left[\varphi - \bar \varphi -\frac12(A + \bar A)\right]\left[\D(\Phi - \frac12(\varphi - \bar \varphi) + \frac14 (A + \bar A))\right]^2\right\}\,. 
\end{aligned}
\end{equation}
This action is invariant under the St\"uckelberg gauge invariance,
 with two independent gauge parameters $\chi$ and $\bar{\chi}$, 
\begin{equation}\label{enhanced_symmetry}
\begin{split}
\delta A &= 2 \chi \,, \qq \delta \bar A = - 2 \bar{\chi} \,, \\
\delta \varphi &= \chi\,, \qquad \;\;\delta \bar \varphi = \bar{\chi}\;. 
\end{split}
\end{equation}
Upon setting 
\begin{equation}
\chi= - \D \zeta \,, \qq \bar \chi= \D \bar \zeta\,, 
\end{equation}
these transformations agree with (the cosmological reduction of) the gauge symmetries found in \cite{Hohm:2016lim}, 
but we observe that the one-dimensional setting is somewhat degenerate in that 
the gauge symmetries  become mere St\"uckelberg transformations. 
Consequently, one may immediately pass over to 
 the gauge invariant quantities
\begin{equation}\label{ApAm}
A^+ \equiv \varphi - \frac12 A \,, 
\qq A^- \equiv \bar \varphi + \frac12 \bar A\,, 
 \end{equation}
obeying $\delta A^{\pm} = 0$, for which the action \eqref{cosmoHSZ_action_infty2} reads 
\begin{equation}
\begin{aligned}
I_{\infty} = \int dt \, n \, &\left\{ - \frac18 \Tr{(\D \Z)^2}  - (\D \Phi)^2 + \frac12 (\D A^+)^2 + \frac12 (\D A^-)^2\right.\\
&\left.+ \, (A^+ - A^-)\left[\D(\Phi - \frac12(A^+ - A^-))\right]^2\right\}\,. 
\end{aligned}
\end{equation}
This manifestly gauge invariant action agrees precisely with \eqref{cosmoHSZ_action_infty1} 
obtained above.  

However, as to be expected, the symmetry \eqref{enhanced_symmetry} does not survive for the full theory \eqref{cosmoHSZ_action_varphis3}. This can be seen by trying to rewrite it in terms of the gauge invariant quantities $A^{\pm}$, \eqref{ApAm}
\begin{equation}\label{cosmoHSZ_action_varphis4}
\begin{aligned}
I = \int dt \, n \, e^{-\Phi + \frac12(A^+ - A^-)} \, &\left\{ \frac{1}{\alpha'}\left[  \frac23 (A^+ + A^-)^3 + \frac12 \Tr{\Z} -\frac16 \Tr{\Z^3}\right]\right.\\
& - \frac18 \Tr{(\D \Z)^2}  - (\D \Phi)^2\\
& + \frac12 (\D A^+)^2 + \frac{1}{\alpha'}\left[\varphi^2 - 4 \varphi A^+ + 2 (A^+)^2\right]\\ 
& + \frac12 (\D A^-)^2 - \frac{1}{\alpha'}\left[\bar \varphi^2 - 4 \bar \varphi A^- + 2 (A^-)^2\right]\\
& \left. + (A^+ - A^-)\left[\D(\Phi - \frac12(A^+ - A^-)\right]^2\right\}\,, 
\end{aligned}
\end{equation}
and observing that the mass-like terms in the third and fourth line explicitly depend on $\varphi$ and $\bar{\varphi}$. 
Curiously, however, \eqref{cosmoHSZ_action_varphis4} is invariant under the non-linear gauge symmetry given by
\begin{equation}
\delta \varphi = (\bar \varphi - 2 A^-)\eta\,, \quad \delta \bar \varphi = (\varphi - 2 A^+ )\eta\,, \quad \delta A^{(\pm)} = 0\,,
\end{equation}
where $\eta(t)$ is an arbitrary gauge parameter.

We finish this analysis by making a final consistency check. We note that in the form \eqref{cosmoHSZ_action_varphis4}, $\varphi$ and $\bar \varphi$ can be integrated out directly since their equations of motion  are given by
\begin{equation}
\varphi = 2 A^+\,, \quad \bar \varphi = 2 A^-\,, 
\end{equation}
and reinserting this into the action yields precisely \eqref{cosmoHSZ_action_massiveApm2}. 

\newpage

\section{Friedmann Universe and tensionless limit}

In this section we analyze the two-derivative equations for FRW backgrounds with a single scale factor. 
To this end we bring the equations into the form of string cosmology with `matter fields', 
which here are the extra fields $A, B$ and $F$. 
While the equations in general are hard to solve, it is straightforward to give the general solutions in the 
tensionless limit $\alpha'\rightarrow \infty$, which  include string frame de Sitter vacua. 
We then outline perturbation theory in $\frac{1}{\alpha'}$ and show how these solutions get corrected. 

\subsection{$\alpha'$-exact Friedmann equations}

We begin by considering the 2-derivative form of the action \eqref{cosmoHSZ_action_2derivative} and expanding $\Z = S + F$, which yields
\begin{equation}\label{I0m}
I = I^{(0)} + I_{m}\,,
\end{equation}
with
\begin{equation}\label{action_0matter}
\begin{aligned}
I^{(0)} &\equiv \int dt \, n \, e^{-\Phi} \, \left\{ - \frac18 \Tr{(\D S)^2} - (\D \Phi)^2 \right\}\,,\\
I_{m} &\equiv \int dt \, n \, e^{-\Phi} \, \left\{ \frac{1}{\alpha'}\left[A B - \frac23 A^3 -\frac12\Tr{S F^2} - \frac16 \Tr{F^3} \right] \right.\\
&\hspace{2.5cm} \left. + \frac14 (\D A)^2 + \frac12 B \D^2 \Phi  - \frac14 \Tr{\D S \D F} - \frac18 \Tr{(\D F)^2}\right\}\,.
\end{aligned}
\end{equation}
In this split $I^{(0)}$ is the standard lowest order, two-derivative gravity action, and $I_m$ encodes the matter content parameterized by $A, B$ and $F$. 
It is important to point out that here we are keeping both projections of the extra fields $F = F_+ + F_-$, and we 
make no a priori assumptions on their dependence on $\alpha'$. 
This is different from the perturbative setup of section \ref{ZtoS}. 
From $I^{(0)}$ one obtains the equations of motion (EOM) for the massless fields, see equation \eqref{EOM0},
while the variation of the matter action is conveniently encoded in the  $O(d,d)$-covariant energy momentum tensor, energy density and dilatonic charge\cite{Gasperini:1991ak, Bernardo:2019bkz} as\footnote{All quantities are defined with a $\sqrt{|g|} e^{\Phi}$ re-scaling compared to standard definitions.}
\begin{subequations}\label{quantities_generic}
	\begin{align}
	\T_M{}^N &\equiv - 2 \frac{e^{\Phi}}{n} S_M{}^P \frac{\delta I_m}{\delta S^P{}_N} \label{TMN} \,,\\
	\rho &\equiv - e^{\Phi} \frac{\delta I_m}{\delta n}\,,\\
	\sigma &\equiv - 2 \frac{e^{\Phi}}{n} \frac{\delta I_m}{\delta \Phi}\,.
	\end{align}
\end{subequations}
The EOM following from  \eqref{I0m} for $S, \Phi$ and $n$ are then given by
\begin{subequations}\label{EOM_matter}
\begin{align}
\left[\Box_{\Phi} S\right]_- S &= - 2 [\T]_-\,, \label{EOMST}\\
2 \D^2 \Phi - (\D \Phi)^2 + \frac18 \Tr{(\D S)^2} &= \frac12 \sigma\,,\\
(\D \Phi)^2 + \frac18 \Tr{(\D S)^2} &= \rho\,,
\end{align}
\end{subequations}
with
\begin{subequations}\label{quantities_HSZ}
	\begin{align}
	[\T]_- &= S \left[- \frac12 \Box_{\Phi} F + \frac{1}{\alpha'} F^2\right]_- \label{Tau}\,,\\
	\rho &= \frac{1}{2 \alpha'} \Tr{S F^2} + \frac{1}{6 \alpha'} \Tr{F^3} - \frac14 \Tr{\D S \D F} - \frac18 \Tr{(\D F)^2} \nn \\
	&\quad - \frac{1}{\alpha'}\left(A B - \frac23 A^3\right) + \frac14 (\D A)^2 + \frac12 B (\D \Phi)^2 -\frac12 \D B \D \Phi \label{rho}\,,\\
	\sigma &= - \frac{1}{\alpha'} \Tr{S F^2} - \frac{1}{3 \alpha'} \Tr{F^3} - \frac12 \Tr{\D S \D F} - \frac14 \Tr{(\D F)^2} \nn \\
	&\quad + \frac{2}{\alpha'}\left(A B - \frac23 A^3\right) + \frac12 (\D A)^2 + 2 B \D^2 \Phi  - B (\D \Phi)^2 + 2 \D B \D \Phi - \D^2 B \label{sigma}\,.
	\end{align}
\end{subequations}
In addition, the EOM for the matter fields are given by
\begin{subequations}\label{EOM_auxiliary}
	\begin{align}
	0 &= \frac14 \Box_{\Phi}(S + F) - \frac{1}{2 \alpha'}\left(S F + F S\right) - \frac{1}{2 \alpha'} F^2 \label{F_+}\,,\\
	A &= - \frac{\alpha'}{2} \D^2 \Phi\,,\\
	B &= 2 A^2 + \frac{\alpha'}{2} \Box_\Phi A = \alpha'{}^2 \left[ - \frac14 \D^4 \Phi + \frac12 (\D^2 \Phi)^2 + \frac14	 \D \Phi \D^3 \Phi\right]\label{B}\,,
	\end{align}
\end{subequations}
where in the last equality of \eqref{B} we used the on-shell value of $A$. 
Indeed, we see here that $A$ and $B$ can be eliminated completely, but we find it convenient to keep them 
in order to be able to work with second-order  equations. 

We also  note that 
reparametrization invariance implies the following 
Noether identity or continuity equation: 
\begin{equation}\label{continuity_equation}
\D \rho + \frac12\Tr{S \D S \T} - \D\Phi(\rho + \frac12 \sigma) = 0\,. 
\end{equation}

Let us now specify to FRW backgrounds, which are characterized by a single scale factor $a(t)$, such that
\begin{equation}\label{S_FLRW}
S = \begin{pmatrix}
0 & a^2 \\
a^{-2} & 0
\end{pmatrix}\,.
\end{equation}
It is also convenient to define the following quantities
\begin{equation}\label{JandK}
\J \equiv \begin{pmatrix}
0 & a^2\\
- a^{-2} & 0
\end{pmatrix}\,, \qquad 
\K \equiv \begin{pmatrix}
1 & 0\\
0 & - 1
\end{pmatrix}\,,
\end{equation}
which, together with $S$, satisfy  identities that will become useful below: 
\begin{equation}\label{SJKidentities}
\begin{aligned}
\J^2 &= - 1\,, \qquad \J = [\J]_-\,,\\ 
\K^2 &= 1\,, \qquad \, \J S = \K\,,\\
\Tr{S} &= \Tr{\K} = \Tr{\J} = 0
 \,. \\
\end{aligned}
\end{equation}
Moreover, for  the Hubble parameter defined as $H(t) \equiv \frac{\D a}{a}$ we have
\begin{equation}\label{LDLidentities}
\D S = 2 H \J\,, \quad \D \J = 2 H S\,.
\end{equation}
Using \eqref{SJKidentities} and \eqref{LDLidentities} we can write out the l.h.s of \eqref{EOM_matter} for the FRW background
\begin{equation}\label{EOM_FLRW_lhs}
\begin{aligned}
\left[\Box_{\Phi} S\right]_- S &\ \rightarrow \  2(\D H - \D \Phi H) \K\,,\\
2 \D^2 \Phi - (\D \Phi)^2 + \frac18 \Tr{(\D S)^2} &  \ \rightarrow 2 \  \D^2 \Phi - (\D \Phi)^2 - d H^2\,,\\
(\D \Phi)^2 + \frac18 \Tr{(\D S)^2} &  \ \rightarrow \  (\D \Phi)^2 - d H^2\,.
\end{aligned}
\end{equation}

For the r.h.s of \eqref{EOM_matter} we need to choose a parameterization for $F$. Instead of considering the extra fields in full generality, however, we will truncate the theory to the subsector $F_- = 0$. Since now we are exploring the theory also in non-perturbative regimes, the argument of section \ref{ZtoS} to fix $F = F_+$ using field redefinitions does not hold anymore. There the theories with or without $F_-$ were equivalent, but here we need to treat $F_- = 0$ as a genuine truncation of the theory.  While we  find  that in general this is not a consistent truncation of the full theory \eqref{action_0matter}, it is a consistent truncation for FRW backgrounds. To see this we split $F = F_+ + F_-$ in \eqref{action_0matter}, compute the EOM for $S, F_{+}$ and $F_-$ separately and then set $F_- = 0$:\footnote{When computing the variation of the action w.r.t. $F_{\pm}$, one should remember that they are constrained fields. As a consequence, part of their variations are determined by $\delta S$ which induce extra terms for the EOM of $S$.} 
\begin{subequations}
\begin{align}
\left. \frac{\delta I}{\delta S} \right|_{F_- = 0} &= \frac18 S \left\{ (S + F_+) \left[\Box_{\Phi}(S + F_+)\right]_- - \left[\Box_{\Phi}(S + F_+)\right]_-(S + F_+) \right\} = 0\,, \label{deltaIS}\\
\left. \frac{\delta I}{\delta F_+} \right|_{F_- = 0} &= \frac14 \left[\Box_{\Phi}(S + F_+)\right]_+ - \frac{1}{\alpha'} S F_+ - \frac{1}{2 \alpha'} F_+^2 = 0\,,\\
\left. \frac{\delta I}{\delta F_-} \right|_{F_- = 0} &= \frac14 \left[\Box_{\Phi}(S + F_+)\right]_- = 0 \label{F-0_constrain}\,.
\end{align}
\end{subequations}
Even though we set $F_-=0$ we still have to satisfy \eqref{F-0_constrain}, which is a constraint on $S$ and $F_+$. While \eqref{F-0_constrain} implies \eqref{deltaIS}, the reverse need not to be true which means that $F_-=0$ is not a consistent truncation in general. In other words, in principle there could be solutions to \eqref{deltaIS} that are not of the form \eqref{F-0_constrain}. For FRW, however, $F_- = 0$ is a consistent truncation because in this case one can show that both equations are equivalent.
Thus, from now on we consider $F_- = 0$ and we proceed to parameterize $F_+ $. The most general ansatz for a +-projected $O(d,d, \mathbb{R})$ tensor consistent with the FRW background \eqref{S_FLRW} is given in terms of two symmetric matrices $\Cp_{m}{}^{n}(t)$ and $\Cm_{m}{}^{n}(t)$, 
\begin{equation}
F_+ = \begin{pmatrix}
\Cm & \Cp a^2\\
a^{-2} \Cp & \Cm
\end{pmatrix}\,,
\end{equation}
where the indices are raised and lowered with $\delta_{m n}$. However, here we 
want to implement  homogeneity and isotropy, as for standard cosmology, by setting 
\begin{equation}\label{truncation}
\Cp_{mn} = \Cp \delta_{mn}\,, \quad \Cm_{mn} = \Cm \delta_{mn} \,, 
\end{equation}
with $f_1$, $f_2$ being functions of time. 
This can be seen to be a consistent truncation 
of the most general ansatz.\footnote{To this end one considers the generic ansatz until getting the final expressions for the EOM \eqref{EOM_matter} and \eqref{EOM_auxiliary} in terms of $\Cp_{m}{}^{n}$ and $\Cm_{m}{}^{n}$ and then verifies  that a diagonal ansatz is consistent.}

In this simplified scheme the matter content is described by just four scalar fields, 
 $A, B, \Cp,$ and $\Cm$. $F_+$ then takes the simple form 
\begin{equation}\label{FCpCm}
F_+ = \begin{pmatrix}
\Cm & 0\\
0 & \Cm
\end{pmatrix} + \begin{pmatrix}
0 & \Cp a^2\\
a^{-2} \Cp & 0
\end{pmatrix} = \Cp S + \Cm {\bf 1}\,,
\end{equation}
where ${\bf 1}$ denotes the unit matrix with components  $\delta_{M}{}^N$. By taking derivatives of \eqref{FCpCm} and using the definition for the generalized energy momentum tensor \eqref{Tau} with $F = F_+$ it follows
\begin{equation}\label{T_deduction}
\begin{aligned}
\left[\T\right]_-  = - \frac12 S \left[\Box_\Phi F_+\right]_-
= \left[ 2 H \D \Cp +  \left(\D H - \D\Phi H \right)\Cp\right] \K \;, 
\end{aligned}
\end{equation}
where we used the matrix defined in (\ref{JandK}). 

In order to give a physical interpretation to \eqref{T_deduction}  notice that for FRW backgrounds \eqref{TMN} we have 
\begin{equation}
\T_M{}^N = - 2 \frac{e^{\Phi}}{n} S_M{}^P \frac{\delta I_m}{\delta S^P{}_N} =
\begin{pmatrix}
- T_m{}^n & 0\\
0 & T^m{}_n
\end{pmatrix}\,,
\end{equation}
with
\begin{equation}
T_{m}{}^n = p\, \delta_m{}^n\,,
\end{equation}
which describes a perfect fluid with pressure $p$. Comparison  with  \eqref{T_deduction} then motivates us to view 
the extra fields of HSZ as describing an effective perfect fluid with an effective pressure determined by $f_{1}$, 
\begin{equation}\label{PCpCm}
\left[\T\right]_- = - p\, \K\,, \quad p = - 2 H \D \Cp - \left(\D H - \D\Phi H \right)\Cp\,.
\end{equation}

For the energy density and the dilatonic charge we just need to insert our expressions for $S, \L, F_+$ and their derivatives in terms of $\Cp$ and $\Cm$ into the definition of $\rho$ and $\sigma$  in \eqref{rho} and \eqref{sigma}. With the help of the identities \eqref{SJKidentities} and \eqref{LDLidentities} one obtains 
\begin{equation}\label{rhosigmaCpCm}
\begin{aligned}
\rho &= \frac{d}{\alpha'}\left( 2 \Cm \Cp + \frac13 \Cm^3 + \Cm \Cp^2 \right) + d H^2\Cp(\Cp + 2) - \frac{d}{4}\left((\D \Cp)^2 + (\D \Cm)^2\right)\\
&\quad - \frac{1}{\alpha'}\left(A B - \frac23 A^3\right) + \frac14 (\D A)^2 + \frac12 B (\D \Phi)^2 -\frac12 \D B \D \Phi\,,\\
\sigma &= - \frac{2 d}{\alpha'}\left( 2 \Cm \Cp + \frac13 \Cm^3 + \Cm \Cp^2 \right) + 2 d H^2\Cp(\Cp + 2) - \frac{d}{2}\left((\D \Cp)^2 + (\D \Cm)^2\right)\\
&\quad + \frac{2}{\alpha'}\left(A B - \frac23 A^3\right) + \frac12 (\D A)^2 + 2 B \D^2 \Phi  - B (\D \Phi)^2 + 2 \D B \D \Phi - \D^2 B\,.\\
\end{aligned}
\end{equation}
With  \eqref{PCpCm} and \eqref{rhosigmaCpCm} the right-hand sides  of the equations of motion 
\eqref{EOM_matter} are completely determined.

Next, for the EOM of the extra fields we can consider just the $+$ projection of \eqref{F_+}, because the minus projection is exactly the EOM of $S$ in \eqref{EOMST} and so it vanishes on-shell. By taking $F_- = 0$, the $+$ projection of \eqref{F_+} reduces to
\begin{equation}
\begin{aligned}
0 &= - \frac{\alpha'}{4} \left[\Box_{\Phi}(S + F_+)\right]_+ + S F_+ + \frac{1}{2} F_+^2\\
&= \left[\frac12\Cp^2 + \frac12 \Cm^2  + \Cp -\frac{\alpha'}{4} \Box_\Phi \Cm\right] {\bf 1} \\
& \quad + \left[(\Cm - \alpha' H^2)(\Cp + 1) -\frac{\alpha'}{4} \Box_\Phi \Cp\right] S\,,
\end{aligned}
\end{equation}
which imply two inequivalent equations (one for each scalar field)
\begin{equation}\label{EOM_CpCm}
\begin{aligned}
\frac12\Cp^2 + \frac12 \Cm^2  + \Cp -\frac{\alpha'}{4} \Box_\Phi \Cm &= 0\,,\\
(\Cm - \alpha' H^2)(\Cp + 1) -\frac{\alpha'}{4} \Box_\Phi \Cp &= 0\,.
\end{aligned}
\end{equation}

When combining \eqref{EOM_FLRW_lhs}, \eqref{PCpCm}, \eqref{rhosigmaCpCm} and \eqref{EOM_CpCm}, we end up with the following non-linear system of coupled second-order differential equations for $a, \Phi, n, A, B, \Cp$ and $\Cm$
\begin{subequations}\label{EOM_FLRW}
	\begin{align}
	p &= \D H - \D \Phi H \label{H}\,,\\
	\frac12(\frac12 \sigma + \rho) &= \D^2 \Phi - d H^2\,, \label{Phi}\\
	\rho &= (\D\Phi)^2 - d H^2\,,\label{n}\\
	0 &= (\Cm - \alpha' H^2)(\Cp + 1) -\frac{\alpha'}{4} \Box_\Phi \Cp\,,\label{f1}\\
	0 &= \frac12\Cp^2 + \frac12 \Cm^2  + \Cp -\frac{\alpha'}{4} \Box_\Phi \Cm\,,\label{f2}\\
	A &= - \frac{\alpha'}{2} \D^2 \Phi\,,\label{A}\\
	B &= 2 A^2 + \frac{\alpha'}{2} \Box_\Phi A \label{BB}\,,
	\end{align}
\end{subequations}
with the `effective matter sources' 
\begin{subequations}\label{prhosigma}
	\begin{align}
	p &= - 2 H \D \Cp - \left(\D H - \D\Phi H \right)\Cp\,,\label{p}\\
	\rho &= \frac{d}{\alpha'}\left( 2 \Cp + \Cp^2 + \frac13 \Cm^2 \right)\Cm + d H^2\Cp(\Cp + 2) - \frac{d}{4}\left((\D \Cp)^2 + (\D \Cm)^2\right) \nn \\
	&\quad - \frac{1}{\alpha'}\left(A B - \frac23 A^3\right) + \frac14 (\D A)^2 + \frac12 B (\D \Phi)^2 -\frac12 \D B \D \Phi\,, \label{rho2}\\
	\frac12(\frac12 \sigma + \rho) &= d H^2\Cp(\Cp + 2) - \frac{d}{4}\left((\D \Cp)^2 + (\D \Cm)^2\right) \nn\\
	&\quad + \frac14 (\D A)^2 + \frac12 B \D^2 \Phi + \frac14 \D B \D \Phi - \frac14 \D^2 B\,.\label{sigma2}
	\end{align}
\end{subequations}
It is instructive to check that these equations are invariant under duality transformations, which in the case of FRW backgrounds reduce to simple full factorized T-dualities
\begin{equation}\label{duality}
a \rightarrow a^{-1}  \quad \Rightarrow \quad H \rightarrow - H\,,
\end{equation}
while the rest of the fields, including $f_1$ and $f_2$, behave as scalars. It is also worth performing some consistency checks before proceeding: the first one comes from the observation that, if $f_1 = \mathcal{O}(\alpha')$ and $f_2 = \mathcal{O}(\alpha')$, the whole system \eqref{EOM_FLRW} reduces to the standard Friedmann equations in vacuum upon neglecting higher orders in $\alpha'$: 
\begin{subequations}\label{EOM_vaccum}
	\begin{align}
	\D H - \D \Phi H &= \mathcal{O}(\alpha')\,,\\
	\D^2 \Phi - d H^2 &= \mathcal{O}(\alpha')\,,\\
	(\D\Phi)^2 - d H^2 &= \mathcal{O}(\alpha')\,.
	\end{align}
\end{subequations}
As a second consistency check we corroborated that the continuity equation \eqref{continuity_equation}, which for FRW backgrounds reads
\begin{equation}
\D \rho + 2 d H p - \D \Phi(\rho + \frac12 \sigma) = 0\,,
\end{equation}
is indeed satisfied for the quantities given in \eqref{prhosigma}.

Remarkably, the above system represents  a non-perturbative and $\alpha'$-complete set of equations for a consistent truncation 
of a theory sharing many features of  genuine string theory. Unfortunately,  finding analytic solutions of \eqref{EOM_FLRW} seems  to be difficult, but as an example of a rather degenerate solution, we can check the one found in \eqref{solutionZ}. 
Specifically, using 
the decomposition $\Z = S + F_+$ and $F_+ = f_1 S + f_2 {\bf 1}$,  the solution $\Z = - {\bf 1}$ corresponds to 
\begin{equation}\label{f1f21}
f_1(t) = f_2(t) = -1\,,
\end{equation}
where $H(t)$ can be arbitrary. While this latter point may be hidden in the way the equations \eqref{EOM_FLRW} are written, one can see that with \eqref{f1f21} all contributions related to the scale factor just disappear. In this case \eqref{f1} and \eqref{f2} are automatically satisfied and \eqref{prhosigma} take the form
\begin{subequations}
	\begin{align}
	p &= \D H - \D \Phi H\,,\\
	\rho &= \frac{2 d}{3 \alpha'} - d H^2 - \frac{1}{\alpha'}\left(A B - \frac23 A^3\right) + \frac14 (\D A)^2 + \frac12 B (\D \Phi)^2 -\frac12 \D B \D \Phi\,,\\
	\frac12(\frac12 \sigma + \rho) &= - d H^2 + \frac14 (\D A)^2 + \frac12 B \D^2 \Phi + \frac14 \D B \D \Phi - \frac14 \D^2 B\,.
	\end{align}
\end{subequations}
By inserting  these results into \eqref{H}, \eqref{Phi} and \eqref{n} it is easy to see that all Hubble parameters cancel out. Finally, by picking the same dilaton value as in \eqref{solutionZ}, 
\begin{equation}
\Phi(t) = \sqrt{\frac{2 d}{3 \alpha'}}\, t + \Phi_0\,,
\end{equation}
with $\Phi_0$ constant, $A=B=0$, the rest of the equations in \eqref{EOM_FLRW} are trivially satisfied.

Apart from this simple case, looking for exact and more complex solutions of the system \eqref{EOM_FLRW} is a complicated task and one should look for simplifications. One option would be to restrict to backgrounds whose matter content describes a barotropic fluid $p = w \rho$ and/or no dilatonic charge, $\sigma = 0$. A second option is to study particular configurations for the dilaton and Hubble parameter and to ask if there exist any configuration of the extra fields $A(t), B(t), f_1(t)$ and $f_2(t)$ such that the equations are satisfied. Another possibility is to study the system perturbatively. In the remainder of this section we will focus  on the last two approaches. For the perturbative computation in particular we will not consider the typical low-energy case of small $\alpha'$, but rather perform an expansion in  $\frac{1}{\alpha'}$ around 
 the tensionless limit $\alpha'\rightarrow \infty$. 

In order to study these two particular paths, we first make some assumptions on the fields and rewrite the system in a more convenient way. We begin by ruling out the somewhat degenerate branch of solutions  by demanding $f_1 \neq -1$. On top of that we also exclude the flat  Minkowski background by demanding $H \neq 0$. From now on we will gauge fix the lapse to $n(t) = 1$ and adopt the following notation for the extra fields
\begin{equation}\label{xy}
x \equiv 1 + f_1 \neq 0\,, \quad y \equiv f_2\,.
\end{equation}
We now observe that under these assumptions equation \eqref{H} can be solved exactly. To see this we begin by noting the simpler expression for the pressure
\begin{equation}\label{pf1}
\begin{aligned}
p &= - 2 H \dot{x} - (\dot{H} - \dot{\Phi} H) (x - 1)\\
&= - 2 H \dot{x} - p (x - 1)\\
&= -2 H \frac{\dot{x}}{x}\\
&=-2 H \partial_t \ln x\,,
\end{aligned}
\end{equation}
where in the second equality we used \eqref{H} and in the third one we isolated $p$ by assuming $x \neq 0$. Plugging \eqref{pf1} into \eqref{H} we get
\begin{equation}
\begin{aligned}
-2 H \partial_t \ln x &= \dot{H} - \dot{\Phi} H\,,\\
\partial_t \ln(x^{-2}) &= \partial_t (\ln H - \Phi)\,,
\end{aligned}
\end{equation}
where in the second line we inverted $H$, which is valid since  we are assuming $H \neq 0$. This equation can be integrated exactly to arrive at
\begin{equation}\label{Hx}
H(t) = Q e^{\Phi(t)} x(t)^{-2}\,, \qquad Q = \text{const.} \neq 0\,.
\end{equation}
This relation tells us that the Hubble parameter is completely determined from the dilaton and one of the extra fields. For the rest of the system we cannot do much without considering particular truncations or certain limits of the theory and so here we just rewrite them in terms of $x$ and $y$: 
\begin{subequations}\label{EOM_FLRWxy}
	\begin{align}
	H &= Q e^{\Phi} x^{-2}\,, 
	 \label{Hxy}  \\
	\ddot{\Phi} &= d H^2 x^2 - \frac{d}{4}\left(\dot{x}^2 + \dot{y}^2\right) + \frac14 (\dot{A})^2 + \frac12 B \ddot{\Phi} + \frac14 \dot{B} \dot{\Phi} - \frac14 \ddot{B}\,, \label{Phixy}\\
	\dot{\Phi}^2&= \frac{d}{\alpha'}\left( x^2 - 1 + \frac13 y^2\right) y + d H^2 x^2 - \frac{d}{4}\left(\dot{x}^2 + \dot{y}^2\right) \nn \\
	&\quad - \frac{1}{\alpha'}\left(A B - \frac23 A^3\right) + \frac14 (\dot{A})^2 + \frac12 B (\dot{ \Phi})^2 -\frac12 \dot{B} \dot{ \Phi}\,,\label{nxy}\\
	0 &= -\frac{\alpha'}{4} \ddot{x} + \frac{\alpha'}{4} \dot{\Phi}\dot{x} + x y - \alpha' H^2 x\,,\label{xxy}\\
	0 &= -\frac{\alpha'}{2}\ddot{y} + \frac{\alpha'}{2}\dot{\Phi}\dot{y} + x^2 + y^2 - 1\,,\label{yxy}\\
	A &= - \frac{\alpha'}{2} \ddot{ \Phi}\,,\label{Axy}\\
	B &= 2 A^2 + \frac{\alpha'}{2} \ddot{A} - \frac{\alpha'}{2} \dot{\Phi}\dot{A} \label{BBxy}\,.
	\end{align}
\end{subequations}

Most of the complexity of the system comes from the terms involving $A$ and $B$. This is because they are the only ones implicitly encoding up to order six in derivatives of the dilaton, as  can be seen by solving \eqref{Axy} and \eqref{BBxy} to express $A$ and $B$ in terms of $\Phi$.  
We did not succeed in finding solutions of the full theory for specific ans\"atze. 
In the following we summarize the backgrounds we studied that turn out not to be exact  solutions:\footnote{We are always considering $x(t) \neq 0$.} 
\begin{itemize}
	\item There are no solutions with constant dilaton $\Phi(t) = \Phi_0 = {\rm const.}$ 
	\item There are no solutions with $H(t) = H_0 = {\rm const.}$ and $\Phi$ being a Laurent polynomial 
	$\Phi(t) = \sum_{k \in \mathbb{Z}} \Phi_k (t - t_0)^k \in \mathbb{R}[t, t^{-1}]$ where all $\Phi_k$ are constant and there are only \textit{finitely-many} non-vanishing of them.
	
	In particular, this result already rules out the possibility of having de Sitter backgrounds in Einstein frame with constant dilaton $\phi(t) = \phi_0 = \text{const}$. To see this we remember that the original dilaton $\phi(t)$ coming from the 2-dimensional string sigma model 
	is related to the duality invariant one $\Phi(t)$ by
	\begin{equation}
	\Phi(t) = 2 \phi(t) - d \ln a(t)\,, \quad \dot{\Phi} = 2 \dot{\phi} - d H\,.
	\end{equation}
	A constant dilaton then imposes the condition
	\begin{equation}\label{constant_dilaton}
	\dot{\Phi} = - d H\,.
	\end{equation}
	On top of that, one can show that for constant dilaton, the Hubble parameter in Einstein frame is constant if and only if $H(t) = H_0 = \text{const.}$ in string frame. In other words, for constant dilaton, de Sitter in string frame is equivalent to de Sitter in Einstein frame. Since \eqref{constant_dilaton} with $H(t) = H_0$ implies a linear dilaton, by ruling out all possible Laurent polynomial forms for $\Phi(t)$, de Sitter solutions in Einstein frame with constant dilaton 
	are also excluded.
	\item We previously observed that by assuming $x = 1 + \mathcal{O}(\alpha')$ and $y = \mathcal{O}(\alpha')$ and performing 
	an $\alpha'$ expansion, the leading order equations of the system \eqref{EOM_FLRWxy} reduce to standard (string) Friedmann equations in vacuum \eqref{EOM_vaccum}. The well-known exact solution to this system is given by
	\begin{equation}\label{solution_vacuum}
	H(t) = \pm \frac{\text{sign}(\omega)}{\sqrt{d}} \frac{1}{(t - t_0)}\,, \quad \Phi(t) = - \log(\omega (t - t_0))\,.
	\end{equation}
	We checked whether this background corresponds to a consistent truncation of the full theory, i.e.~upon including the complete $\alpha'$ dependency. We found that this is only the case if 
	\begin{equation}
	x(t) = 1\,, \quad y(t) = 2 \alpha' \frac{1}{(t - t_0)^2}\,, \quad d = \frac12\,, 
	\end{equation}
with  the final equality of course rendering  this unphysical. 	
\end{itemize}

This analysis shows that some of the simplest backgrounds one can propose for the standard fields $H$ and $\Phi$ are not solutions of the HSZ equations for any configuration of the extra fields $A, B, x$ and $y$. It is worth mentioning, however, that this study just scratches the surface of the whole landscape of possible backgrounds one could propose, and we expect that upon a more exhaustive analyses exact (analytic or numerical) solutions could be found.

\subsection{Tensionless limit, de Sitter solution and $\frac{1}{\alpha'}$ expansion}

While at the present moment we cannot solve the full system \eqref{EOM_FLRWxy} analytically, in the tensionless limit $\alpha' \rightarrow \infty$ the equations are simple enough to obtain the general exact solutions. In this limit, all non-derivative contributions disappear, and \eqref{EOM_FLRWxy} reduces to
\begin{subequations}\label{EOM_tensionless}
	\begin{align}
	H &= Q e^{\Phi} x^{-2}\,, \quad Q = \text{const.} \neq 0 \label{Ht}\,,\\
	\ddot{\Phi} &= d H^2 x^2 - \frac{d}{4}\left(\dot{x}^2 + \dot{y}^2\right) + \frac14 (\dot{A})^2 + \frac12 B \ddot{\Phi} + \frac14 \dot{B} \dot{\Phi} - \frac14 \ddot{B}\,, \label{Phit}\\
	\dot{\Phi}^2&= d H^2 x^2 - \frac{d}{4}\left(\dot{x}^2 + \dot{y}^2\right) + \frac14 (\dot{A})^2 + \frac12 B (\dot{ \Phi})^2 -\frac12 \dot{B} \dot{ \Phi}\,,\label{nt}\\
	0 &= \ddot{x} - \dot{\Phi}\dot{x} + 4 H^2 x\,,\label{xt}\\
	0 &= \ddot{y} - \dot{\Phi}\dot{y} \label{yt}\,,\\
	0 &= \ddot{ \Phi}\,,\label{At}\\
	0 &= \ddot{A} - \dot{\Phi}\dot{A} \label{Bt}\,.
	\end{align}
\end{subequations}
We now turn to \eqref{At}, which implies  a linear dilaton profile:  
\begin{equation}\label{Phitsol}
\Phi(t) = -\omega (t - t_0)\,, 
\end{equation}
where $\omega$ is an integration constant. From now on the solutions are different depending whether $\omega$ vanishes or not. Since the procedure to get both family of solutions is almost identical, we will describe in detail only the $\omega \neq 0$ case and just give the 
final result for vanishing $\omega$. 

Equations \eqref{Bt} and \eqref{yt} take exactly the same form and, upon using \eqref{Phitsol}, they can be solved exactly by
\begin{equation}\label{Aytsol}
\begin{aligned}
A(t) &= A_0 + A_1 e^{-\omega (t - t_0)}\,,\\
y(t) &= y_0 + y_1 e^{-\omega(t - t_0)}\,.
\end{aligned}
\end{equation}
Then, by subtracting \eqref{nt} from \eqref{Phit}, using \eqref{Phitsol} and reordering terms we get a second order differential equation 
for $B$, 
\begin{equation}\label{Beqt}
\ddot{B} + 3 \omega \dot{B} + 2 \omega^2 B - 4 \omega^2 = 0\,,
\end{equation}
which is exactly solved by  
\begin{equation}\label{Btsol}
\begin{aligned}
B(t) &= 2 + B_1 e^{-\omega(t-t_0)} + B_2 e^{-2\omega(t-t_0)}\,.
\end{aligned}
\end{equation}
At this point we have two remaining equations for $x(t)$, namely \eqref{xt} and \eqref{Phit} (or equivalent \eqref{nt}). 
We found it easier to solve \eqref{Phit} because it is a first order differential equation, and then check  \eqref{xt}. 
Inserting \eqref{Ht}, \eqref{Phitsol}, \eqref{Aytsol} and \eqref{Btsol} into \eqref{Phit} we arrive at the first order equation
\begin{equation}\label{xeq}
\dot{x}^2 - \left(4 Q^2 x^{-2} + C_1\right) e^{-2 \omega (t - t_0)} = 0\,, 
\qquad C_1 \equiv \frac{\omega^2}{d}\left(A_1^2 - d y_1^2 - 2B_2 \right)\,,
\end{equation}
where we defined the constant $C_1$ to simplify the notation. By multiplying both sides with $x^2$ and changing variables to $z(t) \equiv x(t)^2$ we arrive at the equation
\begin{equation}\label{zeq}
\dot{z}^2 - \left(16 Q^2 + 4 C_1 z\right) e^{-2 \omega (t - t_0)} = 0\,,
\end{equation}
which has different solutions depending whether $C_1$ vanishes or not, 
\begin{align}\label{zsol}
z(t) &= \pm \frac{4 Q}{\omega} e^{-\omega (t -t_0)} + x_0 \quad \text{if} \quad C_1 = 0\,,\\
z(t) &= \frac{C_1}{\omega^2} \left(e^{-\omega (t -t_0)} + x_0\right)^2 - \frac{4 Q^2}{C_1} \quad \text{if} \quad C_1 \neq 0\,. 
\end{align}
Returning  to the original variable $x(t) = \pm \sqrt{z(t)}$ and plugging the result together with \eqref{Ht} and \eqref{Phitsol} into \eqref{xt} one can verify that the last equation of the system is also satisfied. 

All in all, combining the above results we conclude that, for $\omega \neq 0$, the most general solution to the system \eqref{EOM_tensionless} is given by:
\begin{subequations}\label{tensionless_solution}
	\begin{align}
	\Phi(t) = & - \omega (t - t_0)\,, \quad \omega \neq 0\,,\\
	H(t) = & \ Q e^{- \omega (t - t_0)} x(t)^{-2}\,, \quad Q \neq 0\,,\\
	A(t) = & \ A_0 + A_1 e^{- \omega (t - t_0)}\,,\\
	B(t) = & \ 2 + B_1 e^{- \omega (t - t_0)} + B_2 e^{- 2 \omega (t - t_0)}\,,\\
	y(t) = & \ y_0 + y_1 e^{- \omega (t - t_0)}\,,\\
	x(t) =& \begin{cases}
	\pm 2 \sqrt{ \pm \frac{Q}{\omega} e^{- \omega (t - t_0)} + x_0} &\text{if} \quad C_1 = 0\,,\\
	\pm \sqrt{\frac{C_1}{\omega^2}\left(e^{- \omega (t - t_0)} + x_0\right)^2 - \frac{4 Q^2}{C_1}} &\text{if} \quad C_1 \neq 0\,,
	\end{cases} \ C_1 \equiv \frac{\omega^2}{d}\left(A_1^2 - d y_1^2 - 2B_2 \right)\,.
	\end{align}
\end{subequations}
Repeating identical steps for the $\omega = 0$ case, we get a second set of solutions:
\begin{subequations}\label{tensionless_solutionw0}
	\begin{align}
	\Phi(t) = & \ 0\,,\\
	H(t) = & \ Q x(t)^{-2}\,, \quad Q \neq 0\,,\\
	A(t) = & \ A_0 + A_1 (t - t_0)\,,\\
	B(t) = & \ B_0 + B_1 (t - t_0)\,,\\
	y(t) = & \ y_0 + y_1 (t - t_0)\,,\\
	x(t) = &
	\begin{cases}
	\pm 2 \sqrt{\pm Q (t - t_0) + x_0} &\text{if} \quad C_2 = 0\,,\\
	\pm \sqrt{C_2 [(t - t_0) + x_0]^2 - \frac{4 Q^2}{C_2}}  &\text{if} \quad C_2 \neq 0\,,
	\end{cases} \quad
	C_2 \equiv \frac{1}{d} \left(A_1^2 - d y_1^2\right)\,.
	\end{align}
\end{subequations}
Each of these families is parameterized by several independent free parameters. In particular, for the $\omega \neq 0$ case one can analyze the simplest solution of this  family obtained by taking $A_0 = A_1 = B_1 = B_2 = y_0 = y_1 = x_0 = 0$ and so arriving at 
\begin{subequations}\label{tensionless_solution_dS}
	\begin{align}
	\Phi(t) &= - \omega (t - t_0)\,, \quad H(t) = \text{sgn}(Q) \frac{|\omega|}{4}
	\,, \quad x(t) = \pm 2 \sqrt{\left| \frac{Q}{\omega} \right|} e^{- \frac12 \omega (t - t_0)}\,,\\ A(t) &= y(t) = 0\,, \quad B(t) = 2\,,
	\end{align}
\end{subequations}
where we kept only the real $x(t)$ branch.
Remarkably the Hubble parameter is constant, and hence this solution 
corresponds to a de Sitter background in string frame. Note that the de Sitter scale here is simply an 
integration constant and not determined by a bare parameter in the action, which means that $H$ is fixed by the initial conditions. 
Furthermore, \eqref{tensionless_solution_dS} also admits a de Sitter solution in Einstein frame with constant dilaton for the particular case of $d=4$ (see \eqref{constant_dilaton}), corresponding to five spacetime dimensions. 

The tensionless limit can be interpreted as the zeroth order of a perturbative expansion in small $\frac{1}{\alpha'}$. 
Therefore, in the remainder  of this section we explore the first order correction in $\frac{1}{\alpha'}$ to the system \eqref{EOM_tensionless}. More precisely, we return  to the full system \eqref{EOM_FLRWxy}, write  for all fields 
\begin{equation}\label{Psi01}
\Psi(t) = \Psi^{(0)}(t) + \frac{1}{\alpha'} \Psi^{(1)}(t) + \mathcal{O}\left(\frac{1}{\alpha'{}^2}\right)\,,
\end{equation}
and expand all equations up to first order in $\frac{1}{\alpha'}$. By doing so each equation will split in two, one for each order, the leading one corresponding to the tensionless limit studied in \eqref{EOM_tensionless}. We will not consider corrections to the most general zeroth-order solutions found in \eqref{tensionless_solution} and \eqref{tensionless_solutionw0} but we will restrict to the particular case of \eqref{tensionless_solution_dS} with $Q>0$ and $\omega>0$ for simplicity. However, the following steps should be equally applicable to the general solutions.

Rather than going into each detail, here we show some examples of the procedure described above for the simplest equations. Taking \eqref{Axy} as an example, we expand and keep only up to first order in the string's tension, 
\begin{equation}\label{Aeq1a}
\begin{aligned}
0 &= \ddot{\Phi} + \frac{2}{\alpha'} A\,,\\
0 &= \ddot{\Phi}^{(0)} +\frac{1}{\alpha'}\left(\ddot{\Phi}^{(1)} + 2 A^{(0)}\right) + \mathcal{O}\left(\frac{1}{\alpha'{}^2}\right)\,.
\end{aligned}
\end{equation} 
This splits into two equations, one for the tensionless limit and  a new first order equation that determines $\Phi^{(1)}$ in terms of $A^{(0)}$. Inserting the solution \eqref{tensionless_solution_dS} we see that  \eqref{Aeq1a} is solved by 
\begin{equation}\label{P1}
\Phi^{(1)}(t) = - \omega_1 (t - t_0)\,,
\end{equation}
where $\omega_1$ is a new integration constant, and we omitted a possible constant shift for simplicity. For the second and last explicit calculation, we consider equation \eqref{BBxy}
\begin{equation}\label{Beq1a}
\begin{aligned}
0 &= \ddot{A} - \dot{\Phi} \dot{A} + \frac{1}{\alpha'}\left(4 A^2 - 2 B\right)\,,\\
0 &= \ddot{A}^{(0)} - \dot{\Phi}^{(0)} \dot{A}^{(0)} +  \frac{1}{\alpha'}\left(\ddot{A}^{(1)} - \dot{\Phi}^{(0)} \dot{A}^{(1)} - \dot{\Phi}^{(1)} \dot{A}^{(0)} + 4 (A^{(0)})^2 - 2 B^{(0)}\right) + \mathcal{O}\left(\frac{1}{\alpha'{}^2}\right)\,.
\end{aligned}
\end{equation} 
By inserting \eqref{tensionless_solution} the leading order is automatically solved while the $\frac{1}{\alpha'}$ contribution determines $A^{(1)}$ in terms of the zeroth-order solutions to be
\begin{equation}
A^{(1)}(t) = A_0 + A_1 e^{-\omega (t - t_0)} + \frac{4}{\omega} (t - t_0)\,.
\end{equation}
Following the same procedure, $y^{(1)}(t)$ can be determined by expanding equation \eqref{yxy} and $B^{(1)}(t)$ by expanding the combination of \eqref{Phixy} and \eqref{nxy}. Inserting these and all previous results into the expansion of \eqref{Phixy} we get a first order differential equation for $x^{(1)}(t)$ which can be solved exactly. Finally, at this point all first order corrections were determined, yet we still have to check that the expansion of \eqref{xxy} holds up to first order in the string's tension. We performed all these steps and we found that the extension to the solution \eqref{tensionless_solution_dS} (with positive $Q$ and $\omega$) up to and including first order in $\frac{1}{\alpha'}$ is given by
\begin{subequations}\label{tensionless_solution_dS1}
	\begin{align}
	\Phi(t) &= - \left(\omega_0 + \frac{1}{\alpha'} \omega_1\right) (t - t_0)\,,\\
	H(t) &= \frac{1}{4}\left(\omega_0 + \frac{1}{\alpha'} \omega_1\right) + \frac{1}{\alpha'}\left[\frac{B_2 \omega_0^2}{8 d Q} e^{-\omega_0(t - t_0)} - \frac{x_1}{4} \sqrt{\frac{\omega_0^3}{Q}} e^{\omega_0(t - t_0)}\right]\,,\\
	A(t) &= \frac{1}{\alpha'}\left[A_0 + A_1 e^{-\omega_0(t -t_0)} + \frac{4}{\omega_0} (t - t_0)\right]\,,\\
	B(t) &= 2 + \frac{1}{\alpha'}\left[B_1 e^{-\omega_0(t -t_0)} + B_2 e^{-2 \omega_0(t -t_0)}\right]\,,\\
	y(t) &= \frac{1}{\alpha'}\left[y_0 + y_1 e^{-\omega_0(t -t_0)} - \frac{8 Q}{\omega_0^2} e^{-\omega_0(t -t_0)} (t - t_0) - \frac{2}{\omega_0} (t - t_0)\right]\,,\\
	x(t) &= \pm 2 \sqrt{\frac{Q}{\omega_0}} e^{- \frac12 \omega_0 (t - t_0)}\\
	& \quad \, \pm \frac{1}{\alpha'}e^{-\frac12\omega_0(t -t_0)}\left[x_1 e^{\omega_0(t -t_0)} - \sqrt{\frac{Q}{\omega_0}} \omega_1 \left(\frac{1}{\omega_0} + (t - t_0)\right) - \sqrt{\frac{\omega_0}{Q}} \frac{B_2}{2 d}e^{-\omega_0(t -t_0)}\right]\,,\nn
	\end{align}
\end{subequations}
where we renamed the $\omega$ appearing in \eqref{tensionless_solution_dS} as $\omega_0$. For generic integration constants the solutions \eqref{tensionless_solution_dS1} are not de Sitter vacua. However, there are particular integration constants for  which the zeroth order de Sitter solution \eqref{tensionless_solution_dS} is indeed preserved. For example, by choosing $A_0=A_1=B_1=B_2=y_0=y_1=x_1=0$ and combining the dilaton integration constants into a renormalized $\omega = \omega_0 + \frac{1}{\alpha'} \omega_1$, \eqref{tensionless_solution_dS1} reduces to
\begin{subequations}\label{tensionless_solution_dS2}
	\begin{align}
	\Phi(t) &= - \omega (t - t_0)\,, \quad	H(t) = \frac{\omega}{4}\,, \quad x(t) = \pm 2 \sqrt{\frac{Q}{\omega}} e^{- \frac12 \omega (t - t_0)}\,,\\
	A(t) &= \frac{4}{\alpha' \omega} (t - t_0)\,, \quad B(t) = 2\,,\\
	y(t) &= - \frac{8 Q}{\alpha' \omega^2} e^{-\omega(t -t_0)} (t - t_0) - \frac{2}{\alpha'\omega} (t - t_0)\,, \quad \omega > 0\,, \ Q > 0\,. 
	\end{align}
\end{subequations}
For $\Phi(t), H(t), x(t)$ and $B(t)$ the solutions take  the same structural form as in the tensionless limit \eqref{tensionless_solution_dS}, except that the integration constants are  $\frac{1}{\alpha'}$-corrected. 
It remains as an important open question whether the de Sitter vacua are preserved perturbatively 
at higher orders in $\frac{1}{\alpha'}$.

\section{Conclusions and Outlook}

In this paper we have explored HSZ theory for the purely time-dependent backgrounds of  cosmology. 
While HSZ theory, being based on a non-standard chiral CFT,  is not a conventional string theory, it shares many features 
of string theory such as the presence of a fundamental parameter $\alpha'$ governing higher-derivative corrections 
and exact duality invariance under $O(d,d,\mathbb{R})$. We hope that this theory may thus be a model for how 
to deal with string theory in the regimes we are most  interested in, such as close to the singularities of the big bang 
and black holes. Perhaps most intriguingly, we were able to provide a two-derivative yet $\alpha'$-exact 
reformulation in which the tensionless limit $\alpha'\rightarrow \infty$ can be taken smoothly, and 
we set up perturbation theory with $\frac{1}{\alpha'}$ as the small expansion parameter.

Possible future directions for research include the following: 

\begin{itemize}

\item There are now two known limits of HSZ theory in which a two-derivative reformulation 
is possible: for the quadratic theory in a background field expansion around flat space \cite{Hohm:2016lim}
and for the reduction to one dimension (cosmic time) displayed here. 
It is thus natural to inquire  whether the full HSZ theory admits a reformulation as a 
two-derivative theory. Such a formulation could provide the tensionless limit without any truncations. 

\item In modern cosmology it is equally important to have control not only over the dynamics of the 
homogenous and isotropic backgrounds but also over their fluctuations that generally break the symmetries. 
Recently, the cosmological perturbation theory was explored for a so-called weakly constrained 
double field theory to cubic order in fluctuations, in order to analyze the dynamics of genuine winding modes  \cite{Hohm:2022pfi}. 
It would be interesting to extend this analysis to HSZ theory and, assuming the program of the previous item can be 
accomplished,  to see how the modes behave in the tensionless limit.

\item It is important  to find non-perturbative solutions for the $\alpha'$-complete Friedmann equations, 
ideally  upon adding genuine matter, in order to find semi-realistic string cosmology models. 
In particular, one should  further explore the de Sitter solutions that we found here in the tensionless limit, 
for instance in relation to their stability, as recently explored in a related context in  \cite{Bieniek:2022mrv}.

\item While HSZ theory is not a genuine string theory one might hope that 
the $\alpha'$-complete formulation introduced here could serve as a model for a 
genuine string theory, with the massive string modes playing the role of the extra fields 
employed here. Although genuine string theories feature an infinite number of massive string modes, 
one might still hope that there are formulations not too  dissimilar to the one discussed in this paper.

\item Even if genuine string theories do not allow for such reformulations one may wonder whether
there are other, simpler UV completions of gravity, that display features of string theory, yet lead to a 
version of `string cosmology' that is as manageable as the model explored here.

\end{itemize}

\subsection*{Acknowledgements} 

We thank Robert Brandenberger, Heliudson Bernardo, Roberto Bonezzi, Christoph Chiaffrino, Felipe Diaz-Jaramillo, Allison Pinto, Eric Lescano, Nahuel Miron-Granese 
and Barton Zwiebach   for comments and discussions.

This work  is supported by the European Research Council (ERC) under the European Union's Horizon 2020 research and innovation program (grant agreement No 771862). 
D.~M.~is supported by CONICET. 
T.~C.~is supported by the Deutsche Forschungsgemeinschaft (DFG, German Research Foundation) - Projektnummer 417533893/GRK2575 ``Rethinking Quantum Field Theory". 

\begin{appendix}
\section{Field Redefinitions}

In this appendix we spell out the details of the field redefinitions performed in sec.~3. 
In the action \eqref{Salpha4finalL} we have integrated out the extra  fields, leaving an effective action
purely in terms of the  fields $S, \Phi$ and $n$. In \cite{Hohm:2019jgu} it was  shown  that any action can be brought to a minimal form by performing field redefinitions. In this scheme, there are no higher derivatives of $S$, nor any appearance of $\D \Phi$ or $\Tr{\D S \D S}$ at any higher  order in $\alpha'$. We call any term \textit{removable} if it contains powers or derivatives of at least one of the following objects: $\{\D \L,\D \Phi, \Tr{\L^2} \}$. Inside a Lagrangian these will be denoted as $\Ld$, while for removable terms in the equations of motion  and field redefinitions we will use $\E$ and $\Dd$, respectively. The remaining unremovable couplings are composed of traces of single powers of $\L$.

The procedure to `clean' removable terms order by order was explained in detail in \cite{Hohm:2019jgu}, and here we summarize  the main steps  that are needed. The method relies on field redefinitions  of the effective action \eqref{Salpha4finalL}, 
for which we need the general variation 
\begin{equation}\label{generalVARaction}
\delta I = \int d t \, n \, e^{- \Phi} \left[ \Tr{E_S \delta S} + E_n \frac{\delta n}{n} + E_\Phi \delta \Phi\right]\,, 
\end{equation}
where $E_S $ is the projected version of the original variation, namely $E_S = [\frac{\delta I}{\delta S}]_-$. The equations of motion have an $\alpha'$ expansion, which in our case contains no $\alpha'$ order because $I^{(1)} = 0$, 
\begin{equation}\label{EOM}
E_\Psi = E^{(0)}_\Psi + \alpha'{}^2 E^{(2)}_\Psi + \alpha'{}^3 E^{(3)}_\Psi + \alpha'{}^4 E^{(4)}_\Psi + \mathcal{O}(\alpha'{}^5) = 0 \;. 
\end{equation}
Here we denoted all fields collectively as $\Psi = \{S, \Phi, n\}$. The highest orders in our case $E_\Psi^{(3)}$ and $E_\Psi^{(4)}$ will not be needed to perform field redefinitions. However, we do need $E_\Psi^{(0)}$ and $E_\Psi^{(2)}$, 
given by 
\begin{equation}\label{EOM0}
\begin{aligned}
E_S^{(0)} &= \frac{1}{4} \D \L S - \frac14 \D \Phi \L S\,, \\
E_n^{(0)} &= (\D \Phi)^2 - \frac18 \Tr{\L^2}\,, \\
E_\Phi^{(0)} &= 2 \D^2 \Phi - (\D \Phi)^2 - \frac18 \Tr{\L^2}\,, 
\end{aligned}
\end{equation}
and for $E_\Psi^{(2)}$ by 
\begin{equation}\label{EOM2}
\begin{aligned}
E_S^{(2)} &= \E_S \,,\\
E_n^{(2)} &= \frac{5}{3.2^7} \Tr{\L^6} + \E_n\,, \\
E_\Phi^{(2)} &= \frac{1}{3.2^7} \Tr{\L^6} + \E_\Phi\,. 
\end{aligned}
\end{equation}
As explained before, we collected all removable terms inside $\E_\Psi = \E_\Psi(\D \L, \D \Phi, \Tr{\L^2})$ whose 
explicit form is not needed. 

Let us now consider a general field redefinition 
\begin{equation}
\Psi \rightarrow  \Psi' = \Psi + \delta \Psi\,, 
\end{equation}
which we take to be perturbative in $\alpha'$, so that we can expand 
\begin{equation}
\delta \Psi = \alpha'{}^2 \delta^{(2)} \Psi + \alpha'{}^3 \delta^{(3)} \Psi + \alpha'{}^4 \delta^{(4)} \Psi + \cdots\,. 
\end{equation}
Here we take an expansion of the redefinitions in terms of powers of $\alpha'$, skipping the first order because there are no such terms to remove from the action. 
The action expands as follows: 
\begin{equation}
I'[\Psi']  \equiv I[\Psi+\delta \Psi] = I[\Psi] + \Delta_1I \cdot \delta \Psi 
+\frac{1}{2} \Delta_2I \cdot (\delta \Psi)^2 + \cdots\,,
\end{equation}
where  we defined
$\Delta_n I \equiv \frac{\delta^nI}{\delta \Psi^n}$, 
the first of which is nothing but the equation of motion, $\Delta_1I\equiv \frac{\delta I}{\delta \Psi} \equiv E_{\Psi}$. Each of these variations should be expanded in $\alpha'$ as well
\begin{equation}
\Delta_nI = \Delta_nI^{(0)} + \alpha'{}^2 \Delta_nI^{(2)} + \alpha'{}^3 \Delta_nI^{(3)} + \alpha'{}^4 \Delta_nI^{(4)}+\cdots\,.
\end{equation}
Plugging this expansion into the redefined action $I'$, we obtain 
\begin{equation}\label{fieldredef}
\begin{split}
	I' = I^{(0)}
	&+ \alpha'{}^2 \Big(I^{(2)} + E_\Psi^{(0)} \cdot \delta^{(2)} \Psi\Big)\\
	&+ \alpha'{}^3\Big(I^{(3)} + E_\Psi^{(0)}  \cdot \delta^{(3)} \Psi  \Big)\\
	&+ \alpha'{}^4\Big(I^{(4)} + E_\Psi^{(2)} \cdot \delta^{(2)} \Psi 
	+\frac{1}{2} \Delta_2I^{(0)} \cdot \big(\delta^{(2)}\Psi\big)^2 + E_\Psi^{(0)}  \cdot \delta^{(4)} \Psi  \Big) + \cdots\,,
\end{split}
\end{equation}
where the order $\alpha'{}^3$ does not receive any induced terms from lower orders as a consequence of not having $I^{(1)}$ in the original action. Moreover, $\delta^{(3)} \Psi$ does not propagate to the next order, which will become useful in the following analysis.
In order to bring the action into canonical form we apply the following recipe: first we pick a particular $\delta^{(2)} \Psi$ to bring the 
action to second order in $\alpha'$ to canonical form. This in turn induces new terms proportional to $E_\Psi^{(2)}$ and $\Delta_2I^{(0)}$ into the action of order $\alpha'{}^4$. For the third order we only need to pick some $\delta^{(3)} \Psi$ that brings $I^{(3)}{}'$ to canonical from, yet this step induces no terms at order $\alpha'{}^4$. The induced terms, together with the original $I^{(4)}$, can be brought to canonical form by picking a suitable $\delta^{(4)} \Psi$. Obviously these steps induce terms to all orders in $\alpha'$ but, since we are neglecting orders higher than $\alpha'{}^4$, the above mentioned effects are the only relevant for our work.

In principle, from the above procedure it seems that we must keep track of the field redefinitions  $\delta^{(2)} \Psi$ and  $\delta^{(4)} \Psi$
which can become rather tedious. However, let us show how we can easily obtain $\delta^{(2)} \Psi$ and how $\delta^{(3)} \Psi$ and $\delta^{(4)} \Psi$ are not needed at all. In order to see this, let us suppose that the action contains a term at second order in $\alpha'$ which multiplies the lowest order equations of motion, i.e., 
\begin{equation}\label{S1example}
I =  \cdots + \alpha'{}^2 E_{\Psi}^{(0)} \cdot X(\Psi) + \cdots \;, 
\end{equation}
where $X(\Psi)$ is an arbitrary function of the fields $\Psi$ with four derivatives and the ellipsis denote the remaining terms in the action. By performing a field redefinition of the form
\begin{equation}
\delta^{(2)}\Psi = -X(\Psi)\;. 
\end{equation}
from (\ref{fieldredef}) we then infer that in the redefined action $I'$ the term in (\ref{S1example}) is replaced by
\begin{equation}\label{S1example2}
\begin{split}
	I' = & \cdots - \alpha'{}^4 E_{\Psi}^{(2)} \cdot X(\Psi) + \alpha'{}^4 \frac{1}{2} \Delta_2I^{(0)} \cdot \big(- X(\Psi) \big)^2 + \cdots
	\;, 
\end{split} 
\end{equation}
where the ellipsis denote the same terms as in  (\ref{S1example}), which are unaffected by the redefinition. By comparing \eqref{S1example} and \eqref{S1example2} we see that we managed to remove a term at order $\alpha'{}^2$ in the original action, at expenses of inducing higher order contributions. In particular, we learn two important lessons: 
\begin{enumerate}
	\item We can simply use the equations of motion (\ref{EOM}) as a replacement rule in the form $E_{\Psi}^{(0)} =  -\alpha'{}^2 E_{\Psi}^{(2)} + \dots$, where higher order terms can be ignored as we are considering orders up to an including $\alpha'{}^4$.
	\item We can get access to the induced $\delta^{(2)} \Psi$ by looking at what is left together with $E_\Psi^{(2)}$ in the induced terms. For instance, in the previous example we see that in \eqref{S1example2} this is exactly $\delta^{(2)} \Psi = - X(\Psi)$.
\end{enumerate}

A similar argument follows at $\alpha'{}^3$ and $\alpha'{}^4$ orders. Therein we should pick suitable $\delta^{(3)} \Psi$ and $\delta^{(4)} \Psi$ respectively to eliminate any term proportional to the EOM. The crucial point here is that we do not need their explicit form, because the induced effects of these field redefinitions will become visible at order $\alpha'{}^5$ and higher. In other words, at order $\alpha'{}^3$ and $\alpha'{}^4$ we can just freely use the zeroth order EOM to eliminate any removable contribution.
  
One can then proceed order-by-order in $\alpha'$ by freely using the equations of motion in the action at each order in $\alpha'$, as long as one keeps track of the induced terms and $\delta^{(2)} \Psi$.
To this end we will use the EOM in the form
\begin{subequations}\label{EOMsimplified}
	\begin{align}
	\D \L &= \D \Phi \L - \alpha'{}^2 4 E_S^{(2)} S\;,  \label{EOMSs}\\
	(\D \Phi)^2 &= \frac18 \Tr{\L^2} - \alpha{}^2 E_n^{(2)} \;,\label{EOMns}\\
	\D^2 \Phi &= \frac{1}{2} (\D \Phi)^2 + \frac{1}{16} \Tr{\L^2} - \alpha'{}^2 \frac12 E_\Phi^{(2)} \;. \label{EOMPhis}
	\end{align}
\end{subequations}

By using these equations in a systematic way, one can bring $I^{(2)}, I^{(3)}$ and $I^{(4)}$ to canonical form where all removable terms were redefined away. This systematic approach was explained several times already in
\cite{Hohm:2019jgu} and \cite{Codina:2021cxh}. We will not repeat these steps in detail here but just summarize the main ideas: 

\begin{enumerate}
	\item As a first step one removes all higher derivatives of $\L$ by using \eqref{EOMSs} as many times as needed.
	\item Using \eqref{EOMns} we remove any higher powers of $\D \Phi$.
	\item At this point we make use of \eqref{EOMPhis} to eliminate higher derivatives of the dilaton. In this step higher powers of $\D \Phi$ and higher derivatives of $\L$ could be generated in which case one has to repeat the first two steps. It could be possible that this third step needs to be iterated a few times. 
	\item Reached this point, the only possible contribution from the dilaton comes from a linear term in $\D \Phi$. This can be eliminated by an integration by parts together with \eqref{EOMSs}.
	\item Finally, in order to eliminate terms with $\Tr{\L^2}$ one has to use \eqref{EOMns} to make a $(\D \Phi)^2$ factor reappear, followed by an integration by parts of the dilaton. This will induce higher derivatives of $\D \Phi$ and $\L$ that can be traded for $(\D \Phi)^2$ and $\Tr{\L^2}$ by using \eqref{EOMSs} and \eqref{EOMPhis}. Finally, one uses \eqref{EOMns} once more to bring everything to exactly the same form as the original term that we started from, but with a different coefficient.
\end{enumerate}

We applied this systematic procedure to the second order action 
\begin{equation}
I^{(2)} = \int dt \, n \, e^{-\Phi} \, \left\{- \frac{1}{3.2^7} \Tr{\L^6} - \frac{1}{2^7} \Tr{\D (\L^2) \D (\L^2)} + \frac{1}{16} (\D^3 \Phi)^2 + \frac{1}{12} (\D^2 \Phi)^3\right\}\,,
\end{equation}
and we managed to redefine away all removable terms at expenses of inducing $\alpha'{}^4$ orders
\begin{equation}\label{clean2}
\begin{aligned}
\alpha'{}^2 I^{(2)} + \alpha'{}^2 E_\Psi^{(0)} \cdot \delta^{(2)} \Psi &=  \alpha'{}^2  \int dt \, n \, e^{-\Phi} \, \left\{-\frac{1}{3.2^7} \Tr{\L^6} \right\}\\
& \quad +  \alpha'{}^4 \int dt \, n \, e^{-\Phi} \, \left\{\frac{1}{2^8} \left(E_n^{(2)} - E_\Phi^{(2)}\right) \Tr{\L^4} + \Ld \right\}\\
&= \alpha'{}^2  \int dt \, n \, e^{-\Phi} \, \left\{-\frac{1}{3.2^7} \Tr{\L^6} \right\}\\
&\quad +  \alpha'{}^4 \int dt \, n \, e^{-\Phi} \, \left\{\frac{1}{3. 2^{13}} \Tr{\L^4} \Tr{\L^6} + \Ld \right\}\,,\\
\end{aligned}
\end{equation}
where we used the the second order EOM \eqref{EOM2} for $n$ and $\Phi$ and packaged all removable terms appearing at order $\alpha'{}^4$ in $\Ld = \Ld(\D \L, \D \Phi, \Tr{\L^2})$. From the first line we can read $\delta^{(2)} \Psi$ as explained in point 2. below equation \eqref{S1example2}. These variations are given by
\begin{equation}\label{deltaPsi}
\begin{aligned}
\delta^{(2)} S &= \Dd_S\,,\\
\frac1n\delta^{(2)} n &= \frac{1}{2^8} \Tr{\L^4} + \Dd_n\,,\\
\delta^{(2)} \Phi &= - \frac{1}{2^8} \Tr{\L^4} + \Dd_\Phi\,.
\end{aligned}
\end{equation}
Where $\Dd_\Psi$ denote removable terms that we do not write explicitly.

For $I^{(3)}$ the situation is much simpler since there are no induced terms to keep track of. Therefore, we can just simply eliminate all removable contributions for free, knowing that behind the scenes we picked a suitable $\delta^{(3)} \Psi$. Moreover, from \eqref{Salpha4finalL} we see that $I^{(3)}$ is purely removable which leads to
\begin{equation}\label{clean3}
\alpha'{}^3 I^{(3)} + \alpha'{}^3 E_\Psi^{(0)} \cdot \delta^{(3)} \Psi =  0\,.
\end{equation}
 
Finally, for the fourth order we need the induced terms coming from the cleaning process at order $\alpha'{}^2$. Together with \eqref{clean2}, we also require the quadratic terms $\Delta_2 I^{(0)} \cdot (\delta^{(2)} \Psi)^2$, which can be computed by taking two consecutive variations of the first order action and using \eqref{deltaPsi}. However, we can notice that the only non-removable variations we have are $\frac1n \delta^{(2)} n \propto \delta^{(2)} \Phi \propto \Tr{\L^4}$. Then, whatever $\Delta_2 I^{(0)}$ is, it needs to be an $O(d,d)$ plus 1-dimensional diffeo invariant object with 2 derivatives. There are only three possible objects with these characteristics: $\D^2 \Phi, (\D \Phi)^2$ and $\Tr{\L^2}$, all of them removable. As a consequence, without computing it, we know already that $\Delta_2 I^{(0)} \cdot (\delta^{(2)} \Psi)^2$ with $\delta^{(2)} \Psi$ given by \eqref{deltaPsi}, is removable at order $\alpha'{}^4$.
 
Adding to $I^{(4)}$ (given in \eqref{Salpha4finalL}) the induced terms from \eqref{clean2} together with the (removable) $\Delta_2 I^{(0)} \cdot (\delta^{(2)} \Psi)^2$ we get
\begin{equation}
\alpha'{}^4\Big(I^{(4)} + E_\Psi^{(2)} \cdot \delta^{(2)} \Psi 
+\frac{1}{2} \Delta_2I^{(0)} \cdot \big(\delta^{(2)}\Psi\big)^2 \Big) = \alpha'{}^4 \int dt \, n \, e^{-\Phi} \, \left\{\frac{1}{3. 2^{13}} \Tr{\L^4} \Tr{\L^6} + \Ld \right\}\,.
\end{equation}
Finally, we can finish the process of taking the action to minimal form by choosing a particular $\delta^{(4)} \Psi$ (that we do not need to keep track of) to eliminate all removable terms at order $\alpha'{}^4$, $\Ld$. The final result of this systematic procedure is the HSZ action in the cosmological classification (\ref{HSZcosmoclassification}).

\end{appendix}

\bibliography{CosmoHSZ.bib}
\bibliographystyle{unsrt}

\end{document}